\documentclass[%
 reprint,
 amsmath,amssymb,
 aps,
]{revtex4-2}

\usepackage{graphicx}
\usepackage{dcolumn}
\usepackage{bm}

\usepackage{letltxmacro}
\LetLtxMacro{\originaleqref}{\eqref}
\renewcommand{\eqref}{Eq.~\originaleqref}

\begin{document}

\preprint{APS/123-QED}

\title{Playing it safe: information constrains collective betting strategies}

\author{Philipp Fleig}
\affiliation{Department of Cellular Biophysics, Max Planck Institute for Medical Research, Jahnstra\ss e 29, 69120 Heidelberg, Germany}
\affiliation{Department of Physics \& Astronomy, University of Pennsylvania, Philadelphia, PA 19104, USA}

\author{Vijay Balasubramanian}%
\affiliation{Department of Physics \& Astronomy, University of Pennsylvania, Philadelphia, PA 19104, USA}
\affiliation{Santa Fe Institute,  Santa Fe, NM 87501, USA}%

\date{\today}

\begin{abstract}
Every interaction of a living organism with its environment involves the placement of a bet. Armed with partial knowledge about a stochastic world, the organism must decide its next step or near-term strategy, an act that implicitly or explicitly involves the assumption of a  model of the world. Better information about environmental statistics can improve the bet quality, but in practice resources for information gathering are always limited. We argue that theories of optimal inference dictate that ``complex'' models are harder to infer with bounded information and lead to larger prediction errors.   Thus, we propose a principle of {\it playing it safe} where, given finite information gathering capacity, biological systems should be biased towards simpler models of the world, and thereby to less risky betting strategies. In the framework of Bayesian inference, we show that  there is an optimally safe adaptation strategy determined by the Bayesian prior.  
We then demonstrate that, in the context of stochastic phenotypic switching by bacteria,  implementation of our principle of “playing it safe” increases  fitness (population growth rate) of the bacterial collective. We suggest that the principle applies broadly to problems of adaptation, learning and evolution, and illuminates the types of environments in which organisms are able to thrive.
\end{abstract}

\maketitle


\section*{Introduction}

Risk is an inherent part of life. Whether in pathogen detection~\cite{Burnet}, phenotype selection \cite{Kussell2075,Xue12745}, biochemical and evolutionary mechanisms~\cite{proof_reading,martincorena2012evidence}, foraging and exploration-exploitation strategies~\cite{bees,life_histories,sutton2018reinforcement,vergassola2007infotaxis}, or reproduction~\cite{Cohen_1966,cooperative_breeding}, biological functions are partly shaped by the need to reduce risk.
Broadly, risk arises from  uncertain interactions of 
an organism with the world around it~\cite{Levins_1962}. First, both the environment and the typical sensory apparatus are intrinsically noisy. Thus an organism only has probabilistic information about the state of the world.  Second, any finite system can only observe  some aspects of the world, while others, perhaps most, remain hidden from observation. Thus, an organism inherently functions with partial information, bounding its capacity for rational decision-making. In other words, living systems must necessarily play betting games, producing responses, e.g. expressing a phenotype, detecting an odor, showing an immune response, that lead to outcomes that will probably be beneficial given the limited and uncertain information available for decision-making.

Every betting game has an associated risk. To improve the betting strategy, it is vital to know the odds of the game. An organism can mitigate the risk of its strategy, and minimize the likelihood of wrong bets,  by gathering information  to assess the odds and adapting to the learned statistics of the environment.  Such learning and adaptation is particularly important if  rare, but potentially catastrophic, events can occur.

Living systems can adapt in this way to the environment over many timescales, over the course of evolution, in response to environmental cues, and through ongoing life-long learning.  Given the stochastic nature of the environment from the perspective of an agent inhabiting it, essentially all adaptation processes can be viewed as schemes for inferring a probability distribution~\cite{nemenman2012information,perkins2009strategies}. The ability to infer this distribution  is limited by  physical and temporal constraints of information gathering.  In turn, this limits the capacity for adaptation, and thus increases risk for the system. There are several probabilistic inference frameworks that  explicitly or implicitly account for the limitations of information gathering processes guiding  model choice~\cite{Laplace_1814,Keynes_1921,kullback1997information,Jaynes_2003,MaxEnt_I,MaxEnt_II,RISSANEN_1978,Balasubramanian_1997}.
Here we argue that a key idea arising from these frameworks is that ``complex'' models, in a sense defined in information geometry~\cite{Jeffreys_1946,Balasubramanian_1997}, are difficult to infer robustly from limited data, and that a small inaccuracy in inference of such models can lead to large deviations from the optimal strategy.

Because of this, if information-gathering is limited, it can pay to  bias  inference towards simple models. Thus,  we propose a principle of ``playing it safe'', which dictates that given finite information, biological systems should effectively bias themselves towards simpler models. We derive this principle for a large class of probabilistic models using Bayesian probability theory and information geometry, and show how the bias towards simplicity can be tuned by the choice of prior.
We then illustrate the efficacy of the principle in the classic example of Kelly betting~\cite{Kelly_1956,rivoire2011value}, and demonstrate a biological realization in the phenomenon of stochastic phenotypic switching by bacteria.  We conclude with a discussion of the implications of our finding for the types of environments that are ``learnable'', and in which living systems can prosper.

\section*{Results}

\begin{figure*}[t]
\centering
\includegraphics[width=14.4cm]{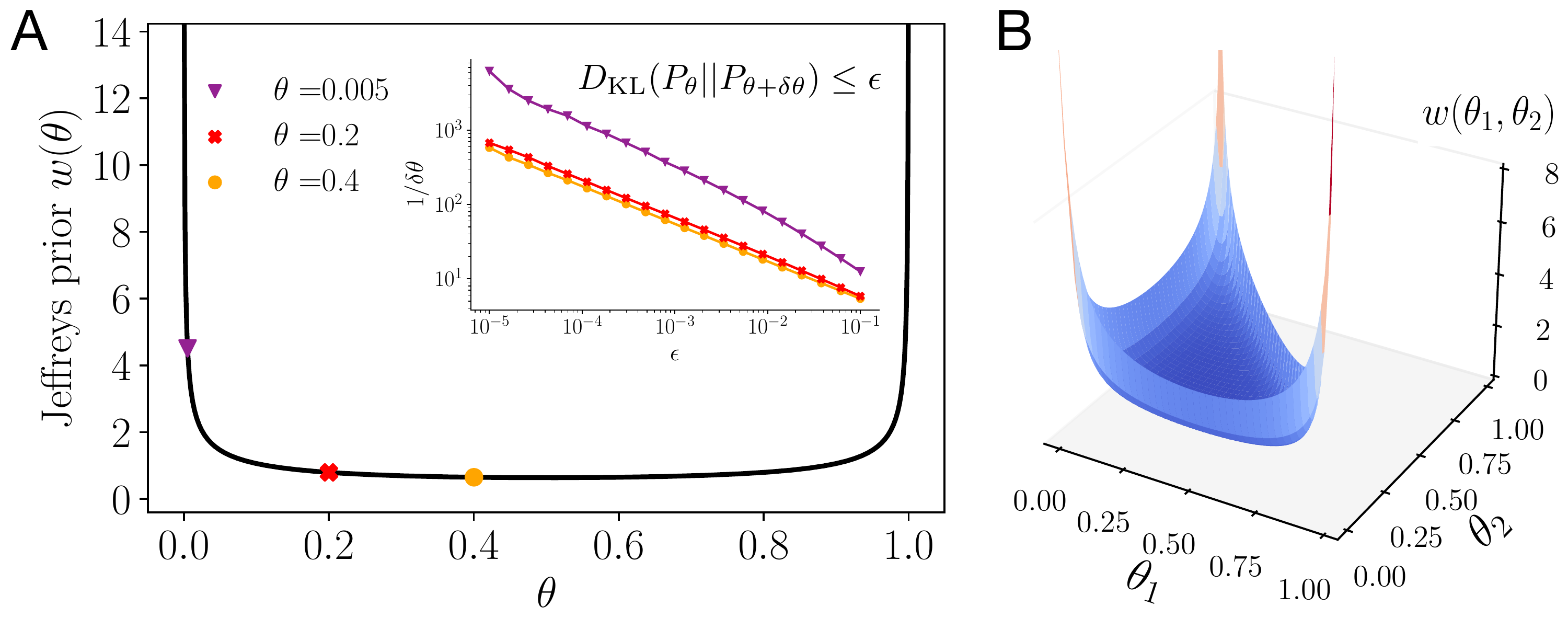}
\caption{Landscape of model complexity dictates the required inference precision. Inferring a model close to a boundary of parameter space requires higher inference precision. (A) Jeffreys prior (black curve) of the Bernoulli model with event probability $\theta$ measures model complexity and diverges at the boundaries of the one-dimensional parameter space. Three example models are indicated by the orange, red, and purple markers. Inset: precision with which the models $\phi=\theta+\delta \theta$ need to be determined to keep the Kullback-Leibler divergence, below some value $\epsilon$. (B) Jeffreys prior for the multinomial model in $d=3$ dimensions. The parameter space is two-dimensional, showing the typical divergences of probabilistic models on boundaries of the space.}\label{fig:fig1}
\end{figure*}

\subsection*{Optimising betting success is hard for complex models}
Betting games are unpredictable by nature. However, bet placement can be optimised by knowing the probabilities controlling events in the game, which often requires estimation of a probability distribution determined by $d$ real parameters, $P(\mathbf x;{\boldsymbol \theta}=\{\theta_1,\ldots,\theta_d\})$,  
denoted here by ${\boldsymbol \theta}$.
In real-world settings, the information required to infer parameters is typically limited. For example, we may have $N$ independent observations $\mathbf x_i$, $\mathbf n=\{\mathbf x_1,\ldots,\mathbf x_N\}$ of a random event which can be used for the inference of parameters. However, finite data leads to statistical fluctuations in the inference process and we need to understand how these affect betting success.

Success in betting games can be quantified through a loss function. In many  games the loss is a function of the game's true probability distribution $P_{\boldsymbol\theta^*}$ and an inferred distribution $Q_{\boldsymbol\phi}$, adopted by the bettor. The true distribution can lie outside the  manifold parameterized by ${\boldsymbol \theta}$ within which the inference is performed, but we will nevertheless use the notation ${\boldsymbol \theta}^*$ for the data generating distribution, because we will often consider cases where this distribution lies within the considered model family. We expect the loss to have a global minimum when ${\boldsymbol \phi}={\boldsymbol \theta}^*$ if the truth lies within the model family under consideration, since exact knowledge of the game's odds allows the bettor to place bets optimally and thus minimise the loss.  If the truth lies outside the model family, the loss should be minimized when ${\boldsymbol \phi}$ comes ``as close as possible'' in terms of the loss function. Here, we adopt the Kullback-Leibler (KL) divergence
\begin{align}\label{eq:DKL}
    \mathcal L_{\boldsymbol\theta^*}(\boldsymbol\phi)\equiv D_\mathrm{KL}(P_{\boldsymbol\theta^*}||Q_{\boldsymbol\phi})=\sum_{\mathbf x} P_{\boldsymbol\theta^*}(\mathbf x)\log \frac{P_{\boldsymbol\theta^*}(\mathbf x)}{Q_{\boldsymbol\phi}(\mathbf x)}
\end{align}
to measure the difference between the true probabilities and the adopted model. The closer the adopted model is to the truth, the smaller the loss. The KL divergence is the loss function of the well-known Kelly betting setup.
An important property of the KL divergence is that for many natural parametrizations of  distributions it diverges on the boundaries of the parameter space.
Changing the parameter by a small amount in such regions leads to large changes in the KL divergence. 

A way to understand the origin of this behaviour is through model complexity.
Consider the Jeffreys prior
\begin{align}
    w({\boldsymbol \theta})=\frac{\sqrt{\det{\mathcal I({\boldsymbol \theta})}}}{\int \mathrm d^d\boldsymbol\psi\sqrt{\det{\mathcal I(\boldsymbol\psi)}}}
\end{align}
where $\mathcal I({\boldsymbol \theta})$ is the Fisher information metric (FIM), obtained by Taylor expansion from the KL divergence between models in the parametric family \cite{Fisher_etal_1922,Amari_1983} (\textit{Materials and Methods}).
In information geometry, this quantity is understood as the  density of distinguishable models, and is regarded as a local measure of complexity in a parametric model family \cite{Amari_1983,Balasubramanian_1997}. 

In regions  where Jeffreys prior is small, i.e., where the model family is {\it simple}, changing a parameter by a small amount does not move the distribution much in model space \cite{Balasubramanian_1997}. However, Jeffreys prior can  become large and even diverge. In such regions, i.e., where the model family is {\it complex}, even models which are close in parameter space are far from each other in model space. In many natural parametrizations, divergences  occur at the boundaries of parameter space (see Figure~\ref{fig:fig1} and \eqref{eq:multinomial_distr} in \textit{Materials and Methods} for the multinomial distribution, and \textit{Appendix}, Section~A for the Poisson and Gaussian distributions).

The multinomial distribution in $d=2$ variables
is described by a single parameter, typically 
the event probability $\theta$.
The corresponding Jeffreys prior in Figure~\ref{fig:fig1}A takes the form $w(\theta)=1/\pi\sqrt{ \theta(1-\theta)}$.
Equivalently $w(\theta) \, \delta\theta^2$ measures the KL divergence between multinomial models parametrized by $\theta$ and $\theta + \delta \theta$.
We see that the density of distinguishable models has a minimum at $\theta=1/2$ and diverges at the boundaries in these particular coordinates.  
The inset in Figure~\ref{fig:fig1}A shows that to keep the KL divergence $D_\mathrm{KL}(P_\theta||P_\phi)$ between two nearby multinomial models below some value $\epsilon$, the parameter $\phi\equiv \theta+\delta \theta$ must lie closer to $\theta$, when $\theta$ itself lies closer to the boundary of the one-dimensional parameter space at $0$. By symmetry, the same holds near the other boundary at $1$. Thus, we see that in this case when the true model is close to a boundary of parameter space, the parameter $\phi$ must be  inferred with higher precision than when the true model lies near a value around $1/2$, in order to achieve the same degree of statistical proximity.

It is hard to optimise the KL divergence in  regions of parameter space that are complex in the sense described above. For example, we may decide to rely on the maximum-likelihood estimate of parameters $\hat{\boldsymbol{\theta}}_\mathrm{ML}$ and define the expected loss as
\begin{align}
    \big<\mathcal L_{\boldsymbol\theta^*}(\hat{\boldsymbol{\theta}}_\mathrm{ML})\big>_{\hat{\boldsymbol\theta}_\mathrm{ML}}\equiv\int\mathrm d^d\hat{\boldsymbol{\theta}}_\mathrm{ML}\, p(\hat{\boldsymbol{\theta}}_\mathrm{ML}) \, \mathcal L_{\boldsymbol\theta^*}(\hat{\boldsymbol{\theta}}_\mathrm{ML})\,,
\end{align}
where $p$ is the distribution of maximum-likelihood values under the true distribution $\boldsymbol\theta^*$ given $N$ observations. In the limit that  $N\rightarrow\infty$, this distribution becomes sharply peaked around $\boldsymbol\theta^*$. However, for finite $N$, $p$ has finite width and the maximum-likelihood estimate $\hat{\boldsymbol{\theta}}_\mathrm{ML}$ is subject to statistical fluctuations due to finite sampling, thus limiting the precision with which the parameters can be determined. In \textit{Appendix}, Section~B we show for the Bernoulli and Poisson models that the KL divergence between the true model and the  expected maximum-likelihood model plus/minus the standard error diverges as the truth approaches the boundary, even if the standard error on the maximum-likelihood estimate decreases to zero with increasing amounts of data. In addition, apart from fluctuations due to finite sampling, in real-world data the observations may additionally be corrupted, effectively introducing an error floor in the empirical estimate.

The question then is what probabilistic model should be selected to control the effects of fluctuations while optimising the placement of bets with finite data?

\subsection*{The playing it safe principle}
We seek a model with the following properties. 
In the large data limit (large $N$), the optimal model has to converge to the maximum-likelihood estimate $\hat{\boldsymbol{\theta}}_\mathrm{ML}$ which itself converges to the true model $\boldsymbol\theta^*$. 
In the opposite limit of small data (small $N$), we should select a model that is less affected by statistical fluctuations. We will parametrize this in terms of a deterministic bias $\boldsymbol\theta_\mathrm{bias}$ and show below how to select it. For finite data size, the optimal model should represent a compromise between the maximum-likelihood model and the bias.
Overall, we make the following Ansatz for the model
\begin{align}\label{eq:Ansatz}
    \boldsymbol\theta = \kappa\,\boldsymbol\theta_\mathrm{bias} + (1-\kappa)\hat{\boldsymbol{\theta}}_\mathrm{ML}\,,
\end{align}
where the parameter $\kappa\in[0,1]$ is used to interpolate between the bias and maximum-likelihood term.
In the next section we show that this Ansatz appears naturally in the inference of a large class of probability distributions (exponential families with a conjugate prior). {\it A priori}, the values of the parameters $\kappa$ and $\boldsymbol\theta_\mathrm{bias}$ can be chosen freely, as long as $\boldsymbol\theta$ lies in the parameter space of the probabilistic model. We determine the optimal choice of these parameters through minimisation of the expected loss, $\big< \mathcal L_{\boldsymbol\theta^*}(\boldsymbol{\theta})\big>_{\hat{\boldsymbol{\theta}}_\mathrm{ML}}$. There are different ways to optimise with respect to these parameters. For instance, we may keep $\kappa$ fixed and optimise for the bias, or vice versa. These lead to different optimal models and which way we pick depends on the setup of the inference we want to solve. We focus on the former approach to optimisation and later discuss the second option in an example.
Thus, we optimise the expected loss with respect to the bias:
\begin{align}
\boldsymbol\theta^\mathrm{opt}_\mathrm{bias}=\underset{\boldsymbol\theta_\mathrm{bias}}{\arg\min} \big< \mathcal L_{\boldsymbol\theta^*}(\boldsymbol{\theta})\big>_{\hat{\boldsymbol{\theta}}_\mathrm{ML}}\,.
\end{align}
To evaluate the condition, it is convenient to write it as a gradient of the expected loss
\begin{align}\label{eq:mincond}
    \frac1\kappa\nabla_{\boldsymbol\theta_\mathrm{bias}}\big< \mathcal L_{\boldsymbol\theta^*}(\boldsymbol{\theta})\big>_{\hat{\boldsymbol{\theta}}_\mathrm{ML}}
    =\nabla_{\boldsymbol\theta}\big< \mathcal L_{\boldsymbol\theta^*}(\boldsymbol{\theta})\big>_{\hat{\boldsymbol{\theta}}_\mathrm{ML}}=0\,.
\end{align}
The gradient field is defined over parameter space and we are looking for optimal points where the gradient vanishes.
As the minima of gradient fields are invariant under coordinate changes, the solution of the optimization problem is invariant to changes in parameterization of the underlying probabilistic model.
However, next we also show that for a large class of probability distributions, there exists a particular parameterization that allows us to understand in detail how the optimum arises from an interplay of the amount of data available about the true model and the geometry of model space. This interplay constrains the regions of parameter space that yield safe betting models, and we thus refer to it as the ``playing it safe" principle.

To make analytic progress, we specialise to exponential families which form a large class of probability distributions, which include many well-known distributions such as the Multinomial, Poisson, and Gaussian distributions etc.~\cite{nielsen2011statistical,wainwright2008graphical}.  An exponential family is a collection of probability densities that can be written in the canonical form
\begin{align}\label{eq:expfamily_density}
    P_{\boldsymbol\eta}(x)=\exp\left(\boldsymbol\eta^T\cdot \mathbf t(x)-F(\boldsymbol\eta)+k(x)\right)\,,
\end{align}
where $\boldsymbol\eta$ is the \emph{canonical parameter} associated with the \textit{sufficient statistic} $\mathbf t(x)$, $F$ is the log-partition function, and $k(x)$ is referred to as the auxiliary carrier term.
Alternatively, we can use the {\it mean parameters} $\boldsymbol\theta$ which are related to the canonical parameters by a one-to-one map (see {\it Materials and Methods}).
The mean parameters often also have a simple relation to commonly used parameterizations of well-known distributions. Examples are the event probabilities in the Multinomial distribution, the average number of counts in the Poisson distribution, and the variance of the Gaussian distribution with zero mean. We give examples of these relations below and in \textit{Appendix}, Sections D and E.

The KL divergence between two densities $P_{\boldsymbol\eta^*}$ and $P_{\boldsymbol\eta}$ of the same exponential family, can be written as a Bregman divergence~\cite{bregman1967relaxation,nielsen2020elementary} between the associated mean parameters:
\begin{align}\label{eq:DKL_Bregman}
    D_\mathrm{KL}(P_{\boldsymbol\eta^*(\boldsymbol{\theta}^*)}|| P_{\boldsymbol\eta(\boldsymbol{\theta})})=B_{F^*}({\boldsymbol\theta^*}:\boldsymbol\theta)\,,
\end{align}
 where $F^*$ is the dual of the log-partition function $F$ (see {\it Materials and Methods}). Bregman divergences form a large class of statistical distances, including many commonly known measures, such as the squared Euclidean distance, the Mahalanobis distance, or the Itakura–Saito distance. We note that in our context, the particular choice of the probabilistic model $P_{\boldsymbol\eta}$, implies a particular Bregman divergence through the dual function $F^*$. 
 In \textit{Appendix}, Section~C1 we show that for exponential families in mean parameterisation, the optimality condition \eqref{eq:mincond} takes the simple form
\begin{align}\label{eq:mincond_exponential_family}
    \big<\mathcal I(\boldsymbol\theta)\delta\boldsymbol\theta\big>_{\hat{\boldsymbol{\theta}}_\mathrm{ML}}=0\,,
\end{align}
which holds even for differences $\delta\boldsymbol\theta\equiv\boldsymbol\theta^*-\boldsymbol\theta$ of finite size. In general, the condition provides $d$ non-linear constraints on the parameters $\kappa$ and $\boldsymbol\theta_\mathrm{bias}$ such that the expected loss is optimised. We give an example illustrating the derivation of this condition for the Bernoulli model in {\it Appendix}, Section~C2.

We now illustrate that this condition, which holds for the mean parameterisation, implies a bias towards models of lower complexity in the sense described in the previous section. For this we focus on the case when the exponential family depends on a single parameter and write out the condition in  \eqref{eq:mincond_exponential_family} explicitly using the definitions of $\delta\theta$ and $\theta$ in \eqref{eq:Ansatz}:
\begin{align}
    \Big<\mathcal I(\theta)\left[\theta^*-\theta\right]\Big>_{\hat{{\theta}}_\mathrm{ML}}&=\nonumber\\
    \Big<\mathcal I(\theta)\left[\theta^*-\hat{{\theta}}_\mathrm{ML}-\kappa(\theta_\mathrm{bias} -\hat{{\theta}}_\mathrm{ML})\right]\Big>_{\hat{{\theta}}_\mathrm{ML}}&=0\,.
\end{align}
\begin{figure}
\centering
\includegraphics[width=8.4cm]{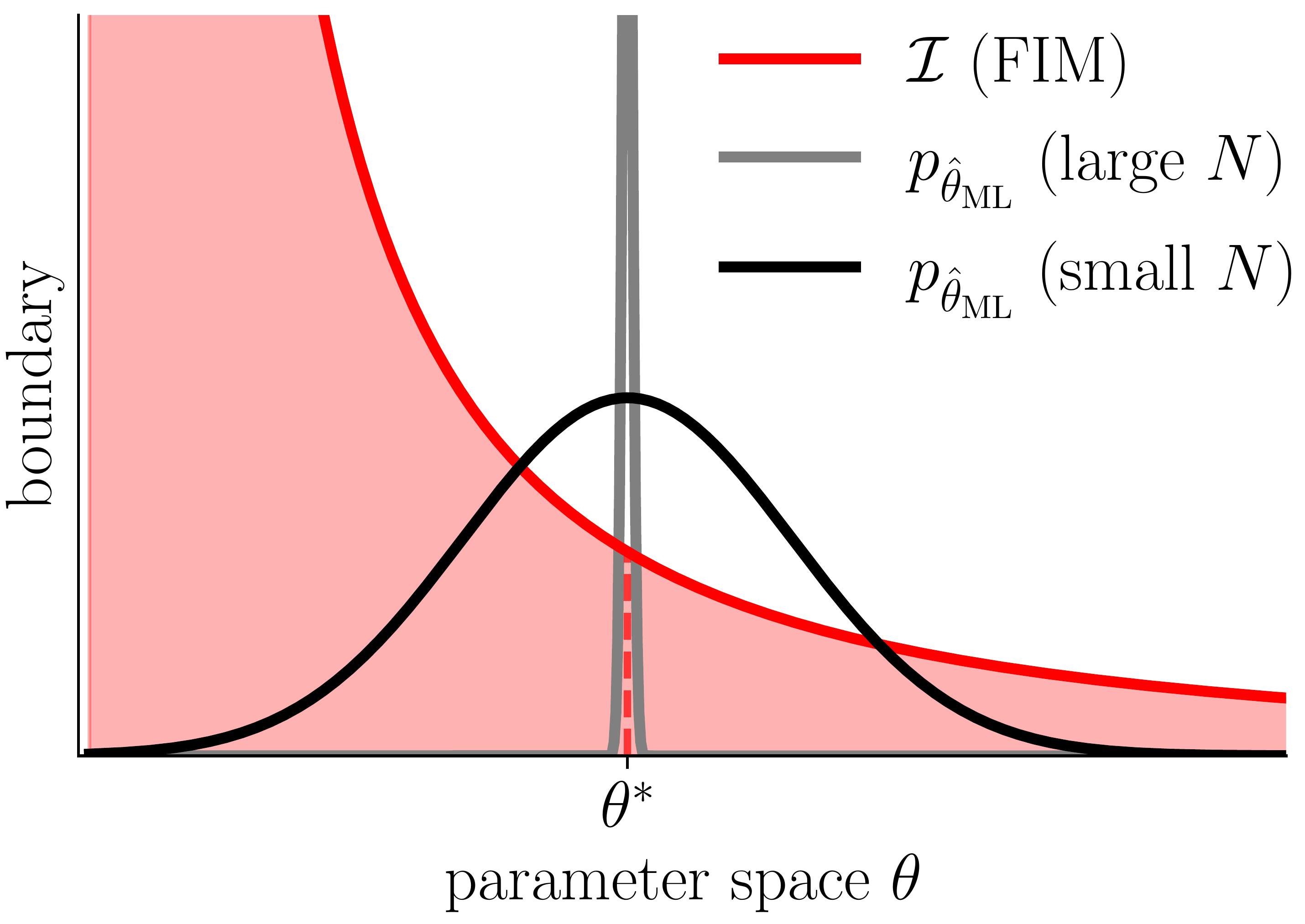}
\caption{Diverging Fisher information at the boundary leads to an asymmetric complexity weight $\mathcal I(\theta)$ around the true model $\theta^*$. We illustrate this for a one-dimensional parameter space with a single boundary. The black and grey curves are the distributions of the empirically observed maximum-likelihood value $\hat\theta_\mathrm{ML}$ for small and large $N$, respectively. For small $N$, the distribution is wide and models to the left of the true model $\theta^*$ (dashed red line), receive a higher complexity weight when evaluating $\langle\mathcal I(\theta)(\theta^*-\theta)\rangle_{\hat\theta_\mathrm{ML}}$ (left-hand side of \eqref{eq:mincond_exponential_family}), than models to the right of the true model.}\label{fig:fig2}
\end{figure}
To illustrate the implications of this condition, we assume an infinite parameter space with a single boundary.
For example, the Poisson distribution with parameter $\lambda\in(0,\infty)$ has such a parameter space. We consider the scenario shown in Figure~\ref{fig:fig2}, where the Fisher information metric $\mathcal I$ diverges to one side of the true model and falls off to small values to the other side.
We then analyse the large and small data limits as illustrated in Figure~\ref{fig:fig2}, and optimise $\theta_\mathrm{bias}$ while $\kappa$ is fixed to a finite value:

Large $N$: The distribution of the maximum-likelihood value becomes sharply peaked around $\theta^*$ and we see that in order to make the bracket inside the expectation value vanish, we require $\theta_\mathrm{bias}\rightarrow\theta^*$.
We note that the local complexity $\mathcal I$ can be arbitrarily large in this limit as it is multiplied by the bracket with value zero. Empirical parameters $\hat{{\theta}}_\mathrm{ML}$ for which $\theta^*-\hat{{\theta}}_\mathrm{ML}$ is not zero are improbable and multiplied by very small probability when taking the expectation value. Thus the entire expectation value will vanish as required.

Small $N$: The distribution of the ML parameter has finite spread around the true model $\theta^*$. For simplicity of argument suppose that the distribution is symmetric around $\theta^*$. To understand the optimization, we go through the possible cases for choosing the bias:
\begin{itemize}
    \item $\theta_\mathrm{bias}=\theta^*$: The expectation value becomes non-zero (either positive or negative), because the ML values on one side of the true model $\theta^*$ receive a higher complexity weight $\mathcal I(\theta)$ than the ML values on the opposite side.
    \item $\theta_\mathrm{bias}$ on the side of $\theta^*$ with diverging complexity: The expectation value assumes even larger (negative or positive) values.
    \item $\theta_\mathrm{bias}$ on the side of $\theta^*$ with diminishing complexity: For the right magnitude of $\theta_\mathrm{bias}$ the expectation value is adjusted to zero, because the models $\theta(\theta_\mathrm{bias}, \hat\theta_\mathrm{ML})$ are biased to receive a smaller complexity weight. This demonstrates the ``playing it safe'' principle for this case.
\end{itemize}
This argument in the one-dimensional case immediately generalises to the case of higher-dimensional parameter spaces. Equation~(\ref{eq:mincond_exponential_family}) provides an independent condition for each of the $d$ dimensions of parameter space and the components of the bias $\theta_\mathrm{bias}$ can thus be adjusted individually to satisfy all conditions.

\subsection*{Optimal Bayesian inference for exponential families}
We now show how the ``playing it safe” principle introduced in the previous section, and in particular the Ansatz  \eqref{eq:Ansatz}, arises
for Bayesian inference of exponential families with  conjugate priors, for which we can provide an analytical treatment. In short, the information contained in the prior determines the function $\kappa$ and the bias $\boldsymbol\theta_\mathrm{bias}$ in \eqref{eq:Ansatz}. We will show that by tuning the conjugate prior through its hyperparameters we can control the strength of the bias relative to the maximum-likelihood term.

We now specialise to exponential families \eqref{eq:expfamily_density} in canonical coordinates $\boldsymbol\eta$. If the true model $\boldsymbol\eta^*$ is unknown, we can at best use the $N$ independent observations $\mathbf x_i$, $\mathbf n=\{\mathbf x_1,\ldots,\mathbf x_N\}$, to determine a Bayesian posterior distribution for ${\boldsymbol \eta}$,
\begin{align}\label{eq:posterior_general}
  P({\boldsymbol \eta}|\mathbf n)=\frac{P(\mathbf n|{\boldsymbol \eta})P({\boldsymbol \eta})}{\int \mathrm d^d\boldsymbol\psi\,P(\mathbf n|\boldsymbol\psi)P(\boldsymbol\psi)}\,,
\end{align}
where $P({\boldsymbol \eta})$ is a prior distribution, and $P(\mathbf n|{\boldsymbol \eta})$ is the likelihood of seeing data ${\mathbf n}$ given the model parameterized by ${\boldsymbol \eta}$. Given the posterior, the goal is to minimise the posterior expected loss
\begin{align}\label{eq:DKL_posterior_loss}
    \big<\mathcal L\big>_\mathrm{post}&=\int\mathrm d\tilde{\boldsymbol\eta}P(\tilde{\boldsymbol\eta}|\mathbf n) D_\mathrm{KL}(Q_{\tilde{\boldsymbol\eta}}|| Q_{\boldsymbol\eta})\\
    &=\int d\tilde{\boldsymbol\theta}\, P(\tilde{\boldsymbol\theta}|\mathbf n)B_{F^*}(\tilde{\boldsymbol\theta}:\boldsymbol\theta)\,,
\end{align}
where in the second line we have used relation \eqref{eq:DKL_Bregman} and 
transformed
variables from canonical to mean parameters also in the posterior density and integral.
We want to determine the minimum value of this expression with respect to $\boldsymbol\theta$
\begin{align}
    \big<\mathcal L\big>_\mathrm{post}(\boldsymbol\theta)=\underset{\boldsymbol\theta}{\min}\big<\mathcal L\big>_\mathrm{post}(\boldsymbol\theta)\,.
\end{align}
This optimal value is also known as the Bregman information.
In \cite{banerjee2005clustering} it was shown that the conditional optimal parameter $\boldsymbol\theta$ is given by the posterior expectation value
\begin{align}
    \boldsymbol\theta=\int d \tilde{\boldsymbol\theta} P(\tilde{\boldsymbol\theta}|\mathbf n)\tilde{\boldsymbol\theta}\,.
\end{align}
Thus, the optimal parameter depends both on the data and the choice of prior. Our next goal is to understand how the prior choice in relation to the available data size affects the optimal parameter. 

Conjugate priors for exponential families allow us to make  general analytic statements in this respect. We denote the conjugate prior of the exponential family by $P(\boldsymbol\eta;\boldsymbol\alpha)$, where $\boldsymbol\alpha$ is a set of hyperparameters.
In \cite{DiaconisYlvisaker_1979} it was shown for exponential families, that under the conjugate prior the posterior mean is given by
\begin{align}                  \boldsymbol\theta(\hat{\boldsymbol\theta}_\mathrm{ML};\boldsymbol\alpha)=\frac{\boldsymbol\alpha+N\hat{\boldsymbol\theta}_\mathrm{ML}}{N_0+N}=\kappa\frac{\boldsymbol\alpha}{N_0}+(1-\kappa)\hat{\boldsymbol\theta}_\mathrm{ML}\,,
\end{align}
where $N_0$ is a function of the hyperparameters $\boldsymbol\alpha$, and $\kappa\equiv N_0/(N_0+N)$. From this expression we can understand the role played by the hyperparameters. $N_0(\boldsymbol \alpha)$ is an threshold number of samples determined by the hyperparameters that sets the scale for whether the actual sample size $N$ is large enough for the evidence to overcome the prior.
Likewise, $\boldsymbol\alpha/N_0$ functions as
effective additional parameters that appear 
on equal footing as the maximum-likelihood parameter. In fact, the quotient $\boldsymbol\alpha/N_0$ is the prior expectation for the parameter $\boldsymbol\theta$, and we denote it by $\boldsymbol\theta_\mathrm{prior}\equiv\mathbb E(\boldsymbol\theta|\boldsymbol\alpha)=\boldsymbol\alpha/N_0$. Overall, this leaves us with the following form of the conditional optimal parameter:
\begin{align}\label{eq:theta_opt}             \boldsymbol\theta=\kappa\boldsymbol\theta_\mathrm{prior}+(1-\kappa)\hat{\boldsymbol\theta}_\mathrm{ML}\,,
\end{align}
with
\begin{align}\label{eq:kappa}
    \kappa=\frac{N_0/N}{N_0/N+1}=\frac{q}{q+1}\,,
\end{align}
where $q\equiv N_0/N$.
We find $\kappa\rightarrow0$ for $N_0/N\rightarrow0$, and $\kappa\rightarrow1$ for $N_0/N\rightarrow\infty$.
Finally, comparison with \eqref{eq:Ansatz} shows that $\boldsymbol\theta_\mathrm{prior}$ takes the role of the bias $\boldsymbol\theta_\mathrm{bias}$, such that overall this confirms our initial Ansatz. 

In the above derivation, the conjugate prior $P(\boldsymbol\eta;\boldsymbol\alpha)$ was defined on the canonical parameters and lead to the linear convex form of \eqref{eq:theta_opt}. However, in the next section and in (\textit{Appendix}, Sections~D and E), we show for the multinomial, Poisson, and Gaussian model, that defining the conjugate prior on the event probabilities, the mean number of counts, and the variance, respectively, equally results in the optimal parameter having convex linear form.
We emphasise that for the development of our theory it is not important on which set of parameters the conjugate prior is defined. What counts for our argument, is that the final expression for the conditional optimal parameter has the linear convex form.

The conditional optimal parameter in convex linear form depends on the data and crucially also on the shape of the prior, set by its hyperparameters. While for a given prior, the loss is minimised by the conditional optimal model, we may further optimise the expected loss with respect to the prior choice which we do by optimising the hyperparameters. Given the convex linear form in \eqref{eq:theta_opt}, we see that there are two natural ways to optimise the hyperparameters. The first, is to make a choice for $\boldsymbol\theta_\mathrm{prior}$, and then optimise the value of $\kappa$ (or, equivalently $q$). The second way is to set the value of $\kappa$ and then optimise $\boldsymbol\theta_\mathrm{prior}$. 
We now illustrate both paths of optimisation for the multinomial distribution.

\subsection*{Optimization for the multinomial model}
\begin{figure*}[t]
\centering
\includegraphics[width=15.4cm]{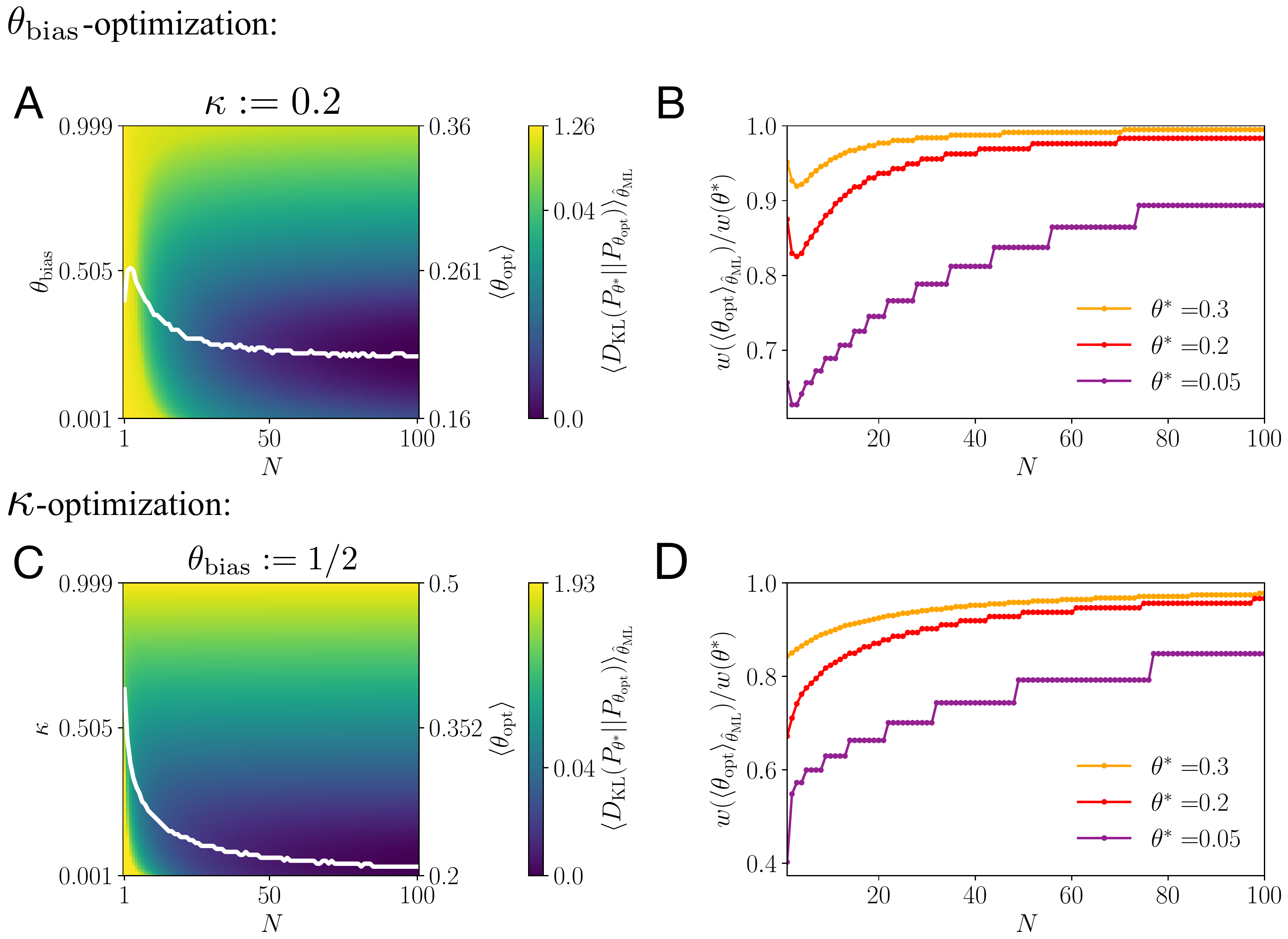}
\caption{Optimization of the expected loss with respect to the prior shows a bias to low complexity models. The true model lies at $\theta^*=0.2$.
(A) The bias $\theta_\mathrm{bias}$ is optimised while $\kappa=0.2$ ($q=0.25$) is kept fixed. The plot shows the landscape of the expected loss as a function of the bias and the number of observations. We are using a non-linear colormap based on the cumulative distribution of the loss. (B) We show the ratio of the local complexity as measured by Jeffreys prior of the expected optimal to that of the true model for three different true models $\theta^*$.
(C) The parameter $\kappa$ is optimised while the bias $\theta_\mathrm{bias}=1/2$ (imposing isotropy on hyperparameter space). The plot shows the landscape of the expected loss as a function of the parameter $\kappa$ and the number of observations. (D) We show the ratio of the local complexity as measured by Jeffreys prior of the expected optimal to that of the true model for three different true models $\theta^*$.}\label{fig:fig3}
\end{figure*}
The multinomial model is the probability distribution of $N$ events, each assuming one of $d$ possible outcomes ({\it Materials and Methods}). The true model is parameterised by the $d$-dimensional set of probabilities ${\boldsymbol \theta}^*$. We note that these probabilities are related to the mean parameters of the multinomial distribution by rescaling the mean parameters with a factor of $1/N$. For simplicity of notation and to emphasize their close relation, we use the same symbol $\boldsymbol\theta$ to denote the probabilities that we used above for the mean parameters. It is natural to collect information on the model by counting event outcomes and computing empirical probabilities $\hat{\boldsymbol \theta}$.
The data $\mathbf n$ consists of the counts of occurrence $x_i$ for each of the $d$ possible outcomes over $N$ observations ($\sum_{i=1}^d x_i=N$). Following our earlier discussion, we use Bayesian inference with a conjugate prior and optimise the conditional model to minimise the expected loss. The likelihood function is the multinomial distribution and its conjugate prior is a Dirichlet distribution given by
\begin{align}
    P(\mathbf n|{\boldsymbol \theta})=\frac{1}{B(\mathbf n)}\prod_{i=1}^d\theta_i^{x_i}\,,\quad
    P({\boldsymbol \theta};\boldsymbol\alpha)=\frac{1}{B(\boldsymbol \alpha)}\prod_{i=1}^d \theta_i^{\alpha_i-1}\,.
\end{align}
The set of $d$ hyperparameters $\alpha_i>0$ determines the prior shape.
Direct calculation yields, as expected, the linear convex form~\eqref{eq:theta_opt} of the conditional optimal model
\begin{align}\label{eq:Laplace}
    \theta_i(\hat{\boldsymbol \theta};\boldsymbol\alpha)=\frac{\frac{\alpha_i}{N}+\hat \theta_i}{\frac{1}{N}\sum_i\alpha_i+1}\,,
\end{align}
where $\hat \theta_i=x_i/N$, are the empirical probabilities. The expression in \eqref{eq:Laplace} is also known as a pseudo-Laplacian law~\cite{Laplace_1814,Lidstone_1920,Nemenman_etal2001} with the $\alpha_i$ being ``pseudocounts". We now specialise to $d=2$, i.e. two possible event outcomes. The conditional optimal model depends on two parameters, $\alpha_1$ and $\alpha_2$ from which we define
\begin{align}\label{eq:multinomial_dequal2_optimal_parameters}
N_0\equiv\alpha_1+\alpha_2\,,\;\boldsymbol\theta_\mathrm{bias}\equiv\frac{\boldsymbol\alpha}{N_0}\,.
\end{align}
The next step is to optimize these hyperparameters in the two possible ways outlined in the previous section.

First, we consider optimization of the bias $\boldsymbol\theta_\mathrm{bias}$, while $\kappa$ is fixed. Since $\theta_2=1-\theta_1$, we are dealing with optimization in one dimension. For simplicity of notation, we suppress component indices in the following. The landscape of the KL loss is shown in Figure~\ref{fig:fig3}A for $\theta^*=0.2$ and $\kappa=0.2$. The curve of the optimal bias choice and the optimized parameter $\langle\theta_\mathrm{opt}\rangle$ (averaged over empirical observations), as a function of $N$ is shown in white. Asymptotically, the curve approaches $\theta^*$ for both parameters. In Figure~\ref{fig:fig3}B, we show how the local complexity of the model with optimal parameter compares to the complexity of the true model as a function of $N$ and for different values of the true model. Asymptotically, the complexity ratio approaches unity, but the closer the true model lies to the boundary, the more data is required to reach the full complexity of the true model. We also observe in Figs.~\ref{fig:fig3}A and B that for small $N$ there is an initial dip in the curves. At first sight this may seem to be in tension with the ``playing it safe” principle. However, this is explained and resolved by the existence of the second boundary of the parameter space of the $d=2$ multinomial model at $\theta=1$. For small $N$, the distribution of maximum-likelihood values is so broad that diverging complexity at the second boundary influences the optimization, pushing the parameter away from the boundary.

Next, we consider optimization of the parameter $\kappa$, while $\theta_\mathrm{bias}$ is held fixed. We make the choice $\alpha_1=\alpha_2$ which, by \eqref{eq:multinomial_dequal2_optimal_parameters} implies $\theta_\mathrm{bias}=1/2$. This is an appealing choice as it implies equal prior knowledge in all directions of parameter space. The landscape of the KL loss is shown in Figure~\ref{fig:fig3}C with $\theta^*=0.2$. The curve of the optimal choice for $\kappa$ and the optimized parameter $\langle\theta_\mathrm{opt}\rangle$ as a function of $N$ is shown in white. Asymptotically, $\kappa$ approaches zero and the optimal model approaches $\theta^*$. In Figure~\ref{fig:fig3}D, we show for different true models, that the complexity ratio approaches unity in the asymptotic limit as expected, but the closer the true model lies to the boundary, the more data is required.

We thus see that either way of optimising implies ``playing it safe” as strategy. In (\textit{Appendix}, Sections~D and E), we demonstrate the optimization of the bias for the Poisson model with unknown average number of counts, and the Gaussian model with unknown variance parameter.

\subsection*{Playing it safe in bacterial phenotypic switching}
\begin{figure*}
\centering
\includegraphics[width=17.4cm]{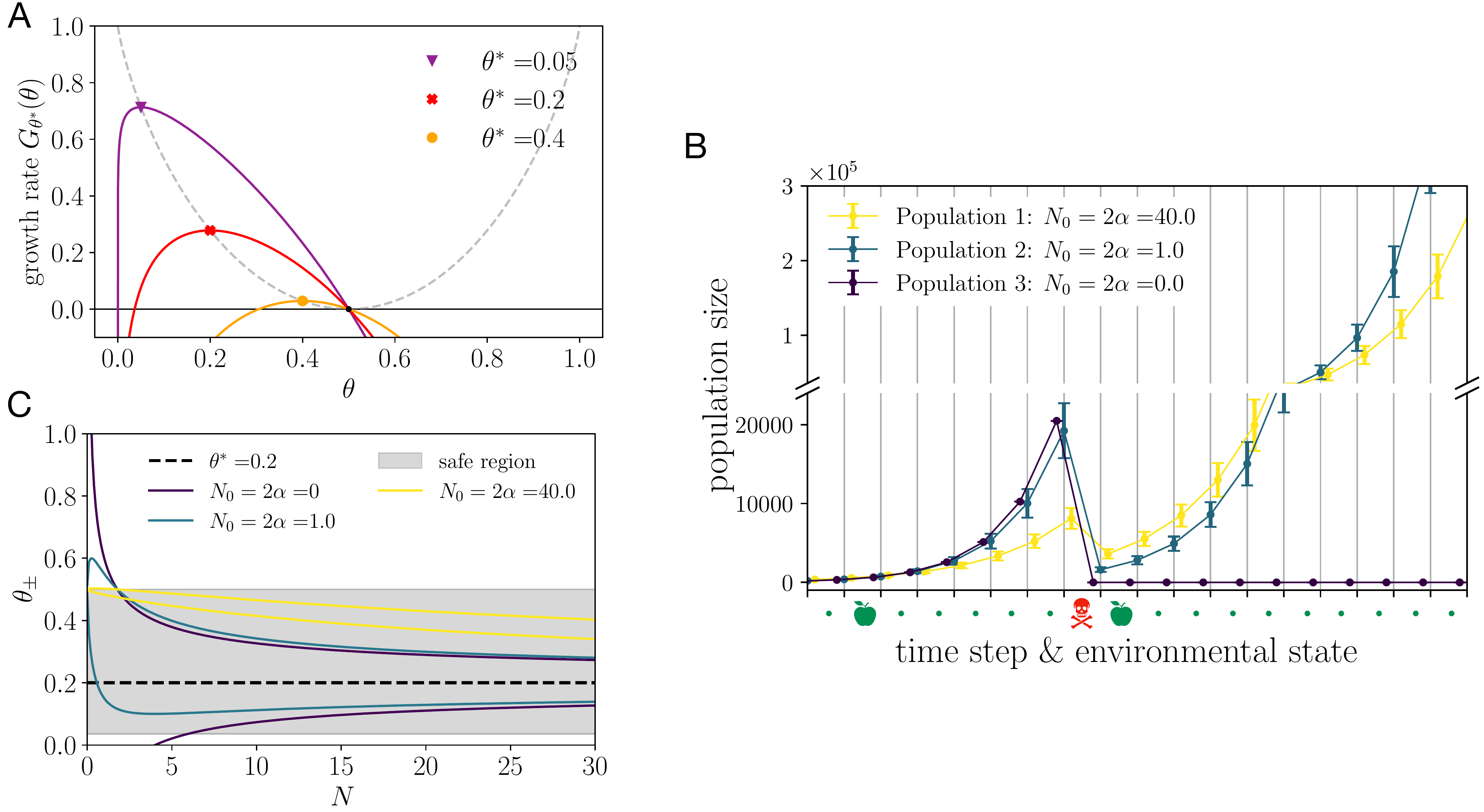}
\caption{Bacteria that place bets according to the ``playing it safe” principle grow more safely. 
(A) Long-term growth rate of the bettor's capital for Kelly betting on independent Bernoulli events as a function of the bettor's assumed model $\theta$ for three values of the true model $\theta^*$ (orange, red, and purple). The maximum is reached when $\theta=\theta^*$.
(B) Three bacterial populations, each of initial size $10$, are exposed to a sequence of binary environmental states (growth or antibiotic) and the development of population sizes is followed. The individuals of each population stochastically switch between a growth and resistant phenotype independently from one and another. All populations keep a memory of the last $N=5$ past environmental states, but each population uses a different learning strategy, set by the hyperparameters $N_0=2\alpha$ and $\theta_\mathrm{bias}=1/2$. The environmental state sequence contains a single antibiotic state (skull), while all other states are growth states (apple). For each population we show the mean population size and the standard deviation. The population with the least safe strategy (dark blue) grows the fastest, but is extinguished by the antibiotic environment. The population with the safest strategy (yellow) grows more slowly, but shows the smallest loss due to antibiotic treatment. The population with intermediate safety (teal), experiences larger loss after antibiotic treatment, but outgrows the population with highest safety. The learning strategy of intermediate safety represents a compromise between too much risk and too much safety. Population curves are computed from fifty independent trials and data points have a small horizontal offset to increase legibility.
(C) The magnitude of fluctuations of the Bayesian conditional optimal model $\theta(\hat \theta;N_0,\theta_\mathrm{bias}=1/2)$ is shown as the plus and minus standard error of the conditional optimal model, $\theta_\pm=\theta(\theta^*\pm\sigma_N;N_0,\theta_\mathrm{bias}=1/2)$, as a function of the number of observations $N$ and the prior choice. Three prior choices from the family of the conjugate prior, set by different values of the hyperparameter $N_0$ (dark blue, teal and yellow) and the fixed bias $\theta_\mathrm{bias}=1/2$. The choice $N_0=0$ corresponds to maximum-likelihood estimation. For large number $N$ of observations, the fluctuations diminish ($q\rightarrow0$ and $\kappa\rightarrow0$) and the conditional optimal model $\theta$ converges to the true model $\theta^*=0.2$ (dotted line). The safe region (gray) constitutes models $\theta$ for which the growth rate function $G_{\theta^*}(\theta)$ is non-negative. For larger values of $N_0$, the standard error of the conditional optimal model $\theta$ is pulled into the safe region, such that a smaller loss in the growth rate is expected.}
\label{fig:fig4}
\end{figure*}
Betting strategies in biology are often found in collective systems where bets can be distributed (or hedged), across the possible outcomes of random events. There are many cases of bet hedging systems in biology~\cite{Mayer5950,Cohen_1966,Gremer_2014};  a particularly well-known example is stochastic phenotypic switching in bacterial populations~\cite{Balaban_etal_2004,Kussell2075}. In the simplest setup, the  population lives in an environment which can assume one of two different states: a nutrient-rich state, allowing the bacterial population to grow, or an antibiotic, hazardous state that kills off bacteria. Each bacterium switches stochastically and independently from the others between a growth and a resistant phenotype. The growth phenotype prospers in the growth state of the environment, but is instantly killed if the environment is in the antibiotic state. Conversely, the resistant phenotype does not grow in the growth environment, but survives antibiotic treatment. It has been shown that in order to maximise  population growth, the bacteria should perform Kelly betting (also known as proportional betting)~\cite{Kelly_1956,Kussell2075}. According to Kelly's theory, the bacteria should stochastically select each phenotype with a probability that is proportional to the probability of occurrence of the corresponding environmental state. We consider the case where the bacteria only have imprecise information about  environmental probabilities, show how the ``playing it safe” principle implies safer betting strategies in this case, and how these choices translate into optimisation of growth of the bacterial population.

In our toy model of stochastic phenotypic switching,  successive discrete environmental states are chosen independently according to a Bernoulli model with a probability $\theta^*$ for the nutrient state, and probability $1-\theta^*$ for the antibiotic states. The bacteria can maximise  population  growth by betting according to Kelly's theory, namely by matching their probabilistic model to the environment. In Kelly betting any mismatch is penalised and the loss is captured by the KL divergence between the true probabilistic model $P_{\theta^*}$ and the model $P_\theta$ assumed by the bacterial population (the bettor). Instead of the loss, it is also instructive to consider the closely related growth rate function derived by Kelly in~\cite{Kelly_1956, rivoire2011value}:
\begin{align}\label{eq:long-term_growthrate}
    G_{\theta^*}(\theta)=[H(P_{1/2})-H(P_{\theta^*})]-D_\mathrm{KL}(P_{\theta^*}||P_\theta)\,,
\end{align}
where $H(\cdot)$ is the entropy.  This function captures the long-term population growth rate, i.e., the  growth rate after a long succession of environmental states. The term in  brackets is the maximal achievable growth rate from which the KL loss is subtracted.
In Figure~\ref{fig:fig4}A we show the long-term growth rate $G_{\theta^*}$ of Kelly betting.
The growth rate has a single maximum when the assumed model is equal to the truth, $\theta=\theta^*$ for which the loss is zero. The further away the adopted model lies from the truth, the smaller the growth rate; it can even become negative if the assumed model lies too far away from the truth.
We will see  (Figures~\ref{fig:fig4}A,B,C and S7) that there is a region of \textit{safe} model choices. This region is given by models $\theta$, for which the long-term population growth rate is non-negative. Models in this region are safe, because while they might not be optimal, they guarantee that there are no long-term losses.

We now assess, for Kelly betting, how the amount of information available about the true model and the prior choice affect the magnitude of fluctuations in the assumed model.
The empirical observations are subject to statistical fluctuations and as discussed above translate to the conditional optimal model. To quantify the magnitude of fluctuations, we consider the {\it standard error} of the empirical estimate $\hat \theta$. For $N$ observations the standard error is given by $\sigma_N=\sqrt{\theta^*(1-\theta^*)/N}$. The fluctuations of the empirical estimate, determine the fluctuations of the conditional optimal model. The magnitude of fluctuations of the conditional optimal strategy \eqref{eq:theta_opt} within one standard deviation, is given by the bounds $\theta_\pm\equiv\theta(\hat\theta^\pm_\mathrm{ML};N_0,\theta_\mathrm{bias})$, where $\hat\theta^\pm_\mathrm{ML}=\theta^*\pm\sigma_N$.

We will examine a toy model of bacterial phenotypic switching, in which environmental states are chosen and phenotypes expressed simultaneously at discrete time steps. The bacterial population then performs collective betting.  We know that with perfect information about the environmental probabilities  the optimal strategy would be to perform proportional betting. This means matching the probability of expressing the growth phenotype with that of the occurrence of the growth state of the environment. However, the bacteria do not know the probability of environmental states. We will suppose that instead they keep an implicit memory of the previous $N$ environmental states, which they use to compute empirical probabilities.

To illustrate the effect of safe learning, we compare the growth of three types of bacterial colonies distinguished only by their learning strategy, as they respond to a sequence of environmental states which include a single antibiotic episode. In Figure~\ref{fig:fig4}B we show how the size of each population develops over time. All bacteria have a memory over the past $N=5$ environmental states. In the extreme case of maximum-likelihood estimation ($N_0=0$, dark blue curve), the population is killed completely by the antibiotic state. In contrast, the safer the learning strategy (larger $N_0$), the smaller the loss a bacterial population suffers due to antibiotic treatment. After the antibiotic treatment, bacteria with the safest strategy recover their numbers most quickly (yellow curve), but they are eventually overtaken by bacteria with less safe learning strategies (teal curve). The observed trade-off between the ability to survive antibiotic treatment and achieve high growth rates is the reason for the existence of an optimally safe learning strategy. The precise choice of optimal strategy depends on the statistics of the environment, as well as potentially other factors such as a future horizon over which safe growth needs to be guaranteed.

In environments which fluctuate more the optimal learning strategy will need to be safer, because the probabilities inferred from a finite memory of past events will have a larger variance.  However, with a maximally safe learning strategy ($\theta = 0.5$), the bacterial population will not be able to achieve any long-term growth, but only keep its population size constant as shown in Figure~\ref{fig:fig4}A. In turn, this means that long-term physical growth is only possible in environments which are not too random. In fact, the more environments are suitable for life to prosper, the closer they are to being deterministic. Near-deterministic environments are thus livable, but as we have shown, rare, yet potentially catastrophic events require organisms to ``play it safe" to survive in even in such worlds.

Finally, in Figure~\ref{fig:fig4}C we contrast different prior choices to show how these affect model safety and plot the standard error bounds of the conditional optimal model $\theta_\pm$ as function of the observations $N$. In this example we fix $\theta_\mathrm{bias}=1/2$ which, by \eqref{eq:multinomial_dequal2_optimal_parameters}, implies $N_0=2\alpha$. 
We observe that a large value of $N_0$ reduces the magnitude of statistical fluctuations and biases the conditional optimal model $\theta_\mathrm{opt}$ towards a probability of $1/2$, i.e. minimal complexity. Overall, the inferred model is pulled into the safe growth region and larger values of $N_0$ thus result in a safer learning strategy. From (\textit{Appendix}, Section~F and Figure~S6) we see that this effect is the more pronounced the closer the true model lies to the boundary of parameter space. Notably, maximum-likelihood estimation ($N_0=2\alpha=0$) is the least safe strategy. For a large number of observations $N$, the fluctuations diminish and the inferred model converges to the true model. However, we observe that safer learning strategies (larger $N_0$) converge to the true model more slowly than less safe strategies. Furthermore, in the limit of maximum safety $q=N_0/N\rightarrow\infty$, the optimal model $\theta_\mathrm{opt}\rightarrow1/2$ for which the growth rate vanishes. This illustrates the trade-off between safe learning and growth maximisation.

\section*{Discussion}
We have proposed that biological systems acting with bounded information should ``play it safe'' by biasing themselves towards models of the environment that are less complex, in the sense of being easier to infer despite statistical fluctuations in limited observed data. We derived this result analytically for exponential families, a very general class of probability distributions, including, e.g., the widely employed multinomial, Poisson, and Gaussian distributions. In the context of collective betting, our results imply that optimal adaptation involves a balance between risk created by uncertainty in inference of the latent distribution, and making full use of the potential for growth or reward.

We illustrated these ideas in the example of bacteria in an environment which mostly supports growth, but occasionally manifests an antibiotic.    In this situation, it pays to maintain diversity between high-growth and low-growth but resistant phenotypes, as a bet-hedging mechanism~\cite{Balaban_etal_2004,Kussell2075}.   Indeed, wild type \textit{Escherichia~coli} show a resistant fraction of $\sim10^{-6}-10^{-5}$~\cite{Moyed768}. We showed how this sort of strategy arises in terms of ``playing it safe'' in an uncertain world.  It will be interesting in the future to apply our analysis to other situations where populations of organisms must make decisions with limited information, for example in bacterial chemotaxis \cite{Mattingly2021.02.22.432091}

We also cast our analysis in the language of Bayesian inference, and noted that the bias towards simpler models can be implemented by an appropriate choice of prior.  The optimal prior depends on knowledge of the true model, or rather its distribution in cases where the generative process in the world itself changes stochastically.    It would be interesting to determine the sorts of evolutionary dynamics that would allow systems to adjust their priors to ``play it safe'' across a given distribution of generative processes.

Another interesting question is to understand how our considerations operate if we use the information maximizing discrete priors discussed in \cite{Mattingly1760,quinn2021information}. We instead used the continuous conjugate priors because it seemed natural that a biological agent might have ways of tuning over a continous family of choices, e.g., by adjusting concentration levels of a protein, or activity levels of a circuit.  By contrast, it is harder to imagine biological methods for selecting over a discrete set of values.   Neverthless, it is important and interesting to consider how our ``playing it safe'' principle might interact with the discrete priors of \cite{Mattingly1760,quinn2021information} because the authors of these articles are precisely concerned with the limitations imposed on inference by having a finite amount of data. Indeed, their discrete priors depend on the number of data points $N$, and are designed to maximize the amount of information gained about the parameters from each observation.  Perhaps there is some way for an organism to incorporate an understanding of the amount of data it is able to accumulate from the past, to effectively discretize its prior on the parameters of the world. It would be useful to understand whether and how some variant of our ``playing it safe'' principle could be realized in that context. Models of biological systems often depend on a very large number of parameters which have been shown to be close to impossible to determine with high precision~\cite{gutenkunst2007universally}. This important observation may suggest that we should define a ``playing it safe'' principle on effective parameters.

Finally, in a time-varying world, inferences are only useful up to some future time horizon. This restricted future utility should limit the value and  precision required of the optimal model, while also increasing the imperative for ``safety'' in light of possible future changes in the world. We can expect trade-offs associated with keeping past memory to play a role in such situations~\cite{schnaack2021risk,lee2022outsourcing}. Perhaps this observation also has a bearing on the widespread observation that Occam's Razor, understood as a preference for simplicity or parsimony, seems to be an organizing principle for mental functions \cite{feldman2016simplicity, koffka2013principles, chater2003simplicity, gershman2013perceptual, tavoni2022human, piasini2023occam}, and that biases towards simple models are helpful in many statistical settings that require inference and prediction \cite{rissanen1996fisher, Balasubramanian_1997,
grunwald2007minimum, bialek2001predictability, Mattingly1760,sachdeva2021optimal}.

\section*{Materials and Methods}

\subsection*{Information geometry basics}
We define the FIM by Taylor expansion of the KL divergence between two parametric distributions $P_{\boldsymbol\theta}$ and $P_{\boldsymbol\phi}$.
The Taylor expansion between two models which are close in parameter space ${\boldsymbol \phi}\equiv{\boldsymbol \theta}^*-\delta{\boldsymbol \theta}$ is given by
\begin{align}
    D_\text{KL}(P_{\boldsymbol \theta^*} ||P_{{\boldsymbol \theta^*} -\delta {\boldsymbol \theta}})=\frac12\mathcal I_{ij}\delta \theta^i\delta \theta^j+\ldots\,,
\end{align}
where summation over repeated indices is implied and the FIM $\mathcal I$ is given by the coefficient of the quadratic term
\begin{align}
    \mathcal I_{ij}({\boldsymbol \theta})=\partial_{\delta \theta^i}\partial_{\delta \theta^j}D_\text{KL}(P_{\boldsymbol \theta} ||P_{{\boldsymbol \theta}- \delta {\boldsymbol \theta}})\Big|_{\delta{\boldsymbol \theta}=\mathbf 0}\,.
\end{align}
As an important example we consider the multinomial distribution
\begin{align}\label{eq:multinomial_distr}
    P(\mathbf x;{\boldsymbol \theta})=\frac{1}{B(\mathbf x)}\prod_{i=1}^d\theta_i^{x_i}\,,
    \text{ with }\;\sum_{i=1}^d\theta_i=1\,,\; x_i\in\{0,...,N\}\,,\\
    \sum_{i=1}^d x_i=N\,, \text{and multivariate Beta } B(\mathbf x)\equiv\frac{\prod_{i=1}^d\Gamma(x_i)}{\sum_{i=1}^d\Gamma(x_i)}\nonumber
\end{align}
where the parameters $\theta_i$ are probabilities and the mean is $N{\boldsymbol \theta}$.
For the multinomial distribution Jeffreys prior takes the form of a Dirichlet distribution
\begin{align}
    w({\boldsymbol \theta})
    &=\frac{\Gamma(d/2)}{\sqrt{\pi}^d}
    \prod_{i=1}^{d}\frac{1}{\sqrt{\theta_i}}\label{eq:sqrtJ}
\end{align}
with $\theta_d = 1 - \sum_{i=1}^{d-1}\theta_i$.
From this form we see that Jeffreys prior has a minimum at $\theta_i=1/d$, but diverges when any of the parameters tend to zero, $\theta_i\rightarrow0$. Analogously, in (\textit{Appendix}, Sections~A) we show for the Poisson and Gaussian distribution, respectively, that Jeffreys prior diverges on the boundaries of the standard parameterization of these distributions. 

\subsection*{Exponential families}
An \textit{exponential family} consists of a collection of densities
\begin{align}
    P_{\boldsymbol\eta}(x)=\exp\left(\boldsymbol\eta^T\cdot \mathbf t(x)-F(\boldsymbol\eta)+k(x)\right)\,,
\end{align}
where $\boldsymbol\eta=[\eta_1\;\eta_2\hdots\;\eta_d]^T$ is the \emph{canonical parameter} associated with the \textit{sufficient statistic} $\mathbf t(x)$, $k(x)$ is the auxiliary carrier term, and $F$ is the log-partition function:
\begin{align}
    F(\boldsymbol\eta)=\log\left(\int \exp(\boldsymbol\eta^T\cdot \mathbf t(x) + k(x))\mathrm d\nu(x)\right)\,,
\end{align}
where $\nu(x)$ the Lebesgue or counting measure. The log-partition function is convex.
If the sufficient statistic is a collection of $d$ functions, the canonical parameter $\boldsymbol\eta$ takes values in the convex set
\begin{align}
    \Omega=\{\boldsymbol\eta\in\mathbb{R}^d\,|\,F(\boldsymbol\eta)<+\infty\}\,.
\end{align}
We consider exponential families where $\Omega$ is an open set, in which case the exponential family is called ${\it regular}$.
Since the log-partition function is a convex function, the set $\Omega$ is also convex. We restrict ourselves to minimal exponential families, which are defined by not having linear constraints amongst the parameters and also not amongst the components of the sufficient statistic. Three examples appear in the table below.
\begin{center}
\begin{tabular}{ c | c | c | c }
 Distribution & $\mathbf t(x)$ & $F(\boldsymbol\eta)$ & $k(x)$ \\ 
\hline
 Multinomial & $[x_1\; x_2 \; \hdots \; x_{d-1}]^T$ & $N\log\left(1+\sum_{i=1}^{d-1}e^{\eta_i}\right)$ & $\log\frac{N!}{ \prod_{i=1}^d x_i!}$ \\ 
 Poisson & $x$ & $e^\eta$ & $ -\log x!$  \\  
 Gaussian & $[x\; x^2]^T$ & $-\frac{\eta_1^2}{4\eta_2}+\frac12\log (-\pi/\eta_2)$ & $0$   
\end{tabular}
\end{center}
An alternative set of parameters is given by the {\it mean parameters} $\boldsymbol\theta$ defined as the expectation of the sufficient statistic
\begin{align}
    \boldsymbol\theta=\mathbb E[\mathbf t(x)]=\nabla_{\boldsymbol\eta} F(\boldsymbol\eta)\,,
\end{align}
where the last relation can be verified by direct calculation. The distribution in mean parameterisation is obtained by inverting this defining equation of the mean parameter and substituting for $\boldsymbol\eta$ in the canonical form of the exponential family.
The analogue of the log-partition function $F$ for the mean parameters is the dual function of $F$ defined by the Legendre-Fenchel transform
\begin{align}
    F^*(\boldsymbol\theta)=\sup_{\boldsymbol\eta \in\Omega}\left\{\boldsymbol\eta^T\cdot \boldsymbol\theta-F(\boldsymbol\eta)\right\}\,.
\end{align}
The dual function $F^*$ can thus be obtained by explicit substitution for $\boldsymbol\eta$ in terms of $\boldsymbol\theta$, after inverting the defining equation of the mean parameter. Finally, we note a relation between $F^*$ and negative entropy $-H(P_{\boldsymbol\eta(\boldsymbol\theta)})$ of the distribution
\begin{align}
    -H(P_{\boldsymbol\eta(\boldsymbol\theta)})&=\int P_{\boldsymbol\eta(\boldsymbol\theta)}\log P_{\boldsymbol\eta(\boldsymbol\theta)} \mathrm{d}\nu(x)\\
    &=\mathbb E[\boldsymbol\eta^T\cdot\mathbf t(x)-F(\boldsymbol\eta)]+\mathbb E[k(x)]\\
    &=F^*(\boldsymbol\theta)+\mathbb E[k(x)]\,,
\end{align}
where we have used the definition of the mean parameters and the defining relation of the dual function.
For distributions which afford a vanishing auxiliary carrier term $k(x)$, the dual function is thus given by the negative entropy. This is for example the case for the Gaussian distribution.


\begin{acknowledgments}
This work was supported in part by the Simons Foundation MMLS Grant 400425 and by the NIH grant R01EB026945. VB thanks the Galileo Galileo Institute where he was a Simons Visiting Scientist, and the Aspen Center for Physics (supported by NSF grant PHY-1607611), for hospitality as this work was completed.
\end{acknowledgments}

\begin{widetext}

\renewcommand\thefigure{S\arabic{figure}}    
\setcounter{figure}{0}  

\appendix

\section{Jeffreys prior for discrete and continuous distributions}
We compute Jeffreys prior for common examples of discrete and continuous distributions. To this end, we define the Fisher information metric (FIM) as the coefficient of the second order term in the Taylor expansion of the Kullback-Leibler (KL) divergence~\eqref{eq:DKL}. If two parameterised distributions, are defined by discrete sets of parameters ${\boldsymbol \theta}=\{\theta_1\,\ldots,\theta_d\}$ and ${\boldsymbol \phi}=\{\phi_1\,\ldots,\phi_d\}$,  respectively, we consider the Taylor expansion of the KL divergence between two models which are close in parameter space ${\boldsymbol \phi}\equiv{\boldsymbol \theta}-\delta{\boldsymbol \theta}$:
\begin{align}
    D_\text{KL}(P_{\boldsymbol\theta} ||P_{\boldsymbol\theta-\delta\boldsymbol\theta})=\frac12{\mathcal I}_{ij}\delta \theta^i\delta \theta^j+\ldots\,,
\end{align}
where summation over repeated indices is implied and $\mathcal I$ is the FIM, given by
\begin{align}
    {\mathcal I}_{ij}({\boldsymbol \theta})=\partial_{\delta \theta^i}\partial_{\delta \theta^j}D_\text{KL}(P_{\boldsymbol\theta} ||P_{\boldsymbol\theta-\delta\boldsymbol\theta})\Big|_{\delta{\boldsymbol \theta}=\mathbf 0}\,.
\end{align}
The normalised Jeffreys prior is computed from the FIM through the relation
\begin{align}
    w({\boldsymbol \theta})=\frac{\sqrt{\det \mathcal I({\boldsymbol \theta})}}{\int\mathrm d^d{\boldsymbol \psi} \sqrt{\det \mathcal I({\boldsymbol \psi})}}\,.
\end{align}
\subsection{Multinomial distribution}
We define the multinomial distributions with parameters ${\boldsymbol \theta}$ and ${\boldsymbol \theta}'$ as:
\begin{align}
    P_{\boldsymbol\theta}(\mathbf x)=\frac{1}{B(\mathbf x)}\prod_{i=1}^d\theta_i^{x_i}\,,\;
    Q_{\boldsymbol\theta'}(\mathbf x)=\frac{1}{B(\mathbf x)}\prod_{i=1}^d{\theta'}_i^{x_i}
\end{align}
with
\begin{align}
\sum_{i=1}^d\theta_i=\sum_{i=1}^d{\theta'}_i=1\,,\; x_i\in\{0,...,N\}\text{ and }\sum_{i=1}^d x_i=N\,.
\end{align}
The KL divergence between the two distributions is given by
\begin{align}
    D_\mathrm{KL}(P_{\boldsymbol\theta} ||Q_{\boldsymbol\theta'})&=\sum_{\mathbf x}\frac{1}{B(\mathbf x)}\prod^d_{i=1}\theta_i^{x_i}\log\left(\frac{\prod_{j}\theta_j^{x_j}}{\prod_j{\theta'}_j^{x_j}}\right)\\
    &=\sum_j(\log\theta_j-\log\theta_j')\left(\sum_\mathbf{x}\frac{x_j}{B(\mathbf x)}\prod_{i=1}^d\theta_i^{x_i}\right)\\
    &=\sum_j(\log\theta_j-\log\theta_j')N\theta_j\,,
\end{align}
where in the last step we have evaluated the mean of the multinomial distribution.
The next step is to Taylor expand the KL divergence around ${\boldsymbol \theta}$, by setting ${\boldsymbol \theta}'={\boldsymbol \theta}-\delta{\boldsymbol \theta}$. For the Taylor expansion we find
\begin{align}
    D_\mathrm{KL}(P_{\boldsymbol\theta}||Q_{\boldsymbol\theta'})&=\sum_j\left(\frac{\delta\theta_j}{\theta_j}+\frac{\delta\theta_j^2}{2\theta_j^2}+O(\delta\theta_j^3)\right)N\theta_j=\sum_j\left(\frac{\delta\theta_j^2}{2\theta_j^2}+O(\delta\theta_j^3)\right)N\theta_j\,,
\end{align}
where in the last step the linear term vanishes since $\sum_i\theta_i=\sum_i\theta_i'=1$ and $\theta_i'=\theta_i-\delta\theta_i$. From the quadratic  term we read off the form of the FIM
\begin{align}
    {\mathcal I}_{ij}({\boldsymbol \theta})=N\frac{\delta_{ij}}{\theta_i}\,,
\end{align}
where $\delta_{ij}$ is the Kronecker delta. Taking the square root of the determinant and computing the normalisation factor finally gives us Jeffreys prior:
\begin{align}
    w({\boldsymbol \theta})=Dir({\boldsymbol \theta};1/2,\ldots,1/2)=\frac{\Gamma(d/2)}{\sqrt{\pi}^d}\frac{1}{\sqrt{1-\sum_{i=1}^{d-1}\theta_i}}\prod_{i=1}^{d-1}\frac{1}{\sqrt{\theta_i}}\,,\label{eq:Multinomial_Jeffreys}
\end{align}
where $\mathrm{Dir}$ is the Dirichlet distribution.

\subsection{Poisson distribution}
We consider two Poisson distributions with parameters ${\lambda'}$ and $\lambda$:
\begin{align}
    P_{\lambda'}(n)=\frac{{\lambda'}^ne^{-{\lambda'}}}{n!}\,,\;
    Q_\lambda(n)=\frac{\lambda^ne^{-\lambda}}{n!}\,.
\end{align}
The KL divergence between the two distributions is given by
\begin{align}
    D_\mathrm{KL}(P_{\lambda'}||Q_\lambda)&=\sum_n\frac{{\lambda'}^ne^{-{\lambda'}}}{n!}\log\left(\frac{{\lambda'}^ne^{-{\lambda'}}}{n!}\frac{n!}{\lambda^ne^{-\lambda}}\right)={\lambda'}\log\frac{{\lambda'}}{\lambda}-{\lambda'}+\lambda\,.
\end{align}
The next step is to Taylor expand the KL divergence around ${\lambda'}$, by setting $\lambda={\lambda'}+\delta\lambda$. For the Taylor expansion we find
\begin{align}
    D_\mathrm{KL}(P_{\lambda'}||Q_\lambda)&=\frac{\delta\lambda^2}{2{\lambda'}}+O(\delta\lambda^3)\,.
\end{align}
From the quadratic term we read off the form of the FIM
\begin{align}
    {\mathcal I}({\lambda'})=\frac{1}{{\lambda'}}\,.
\end{align}
Taking the square root  gives us Jeffreys prior:
\begin{align}
    w({\lambda'})=\frac{1}{\sqrt{{\lambda'}}}\,,\label{eq:Poisson_Jeffreys}
\end{align}
Jeffreys prior is shown in Figure \ref{fig:S4}. It diverges on the boundary of parameter space for ${\lambda'}\rightarrow0$ and approach zero as for ${\lambda'}\rightarrow\infty$.
\begin{figure}
\centering
\includegraphics[width=11.4cm]{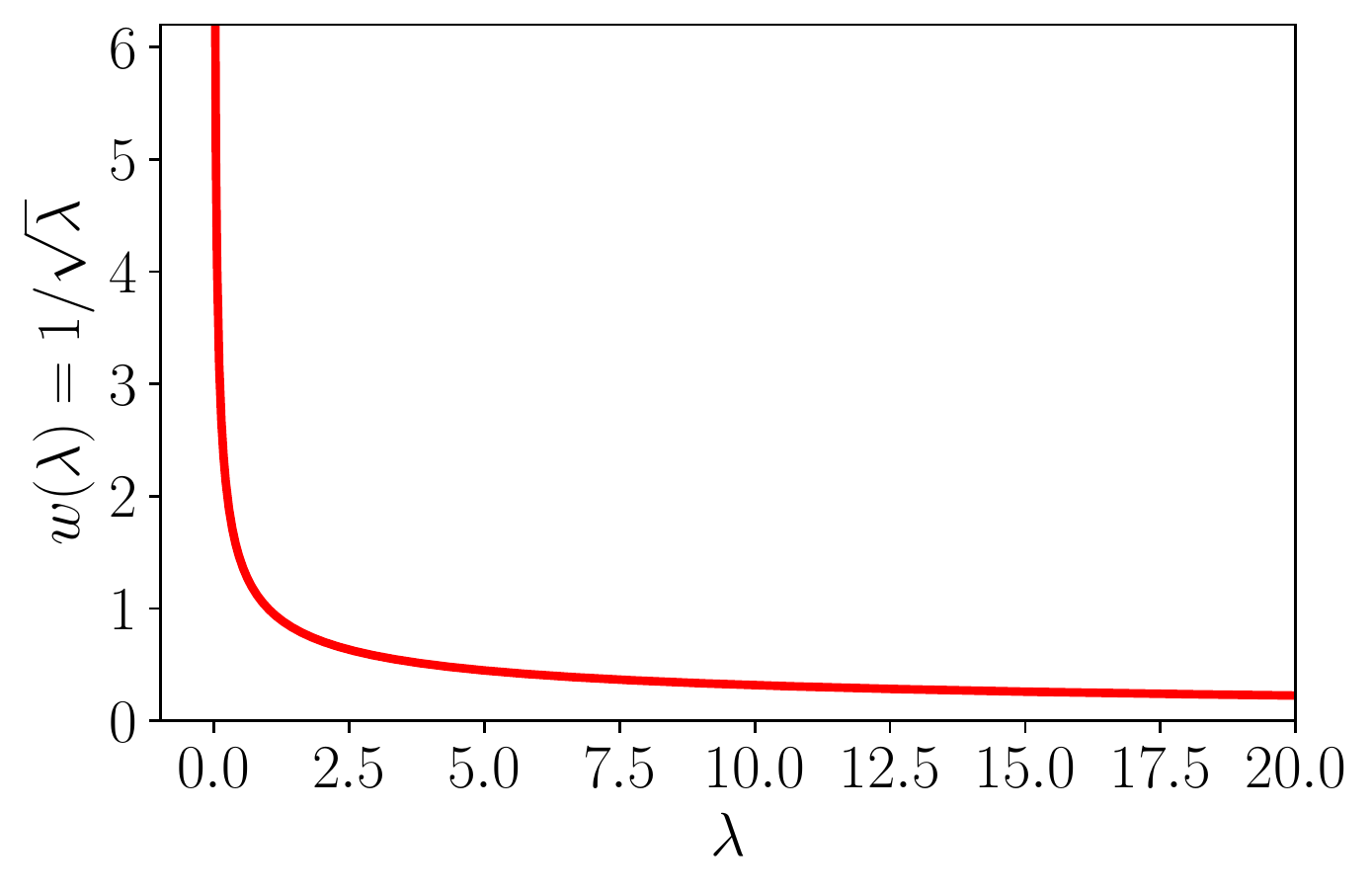}
\caption{Jeffreys prior for the Poisson distribution with rate $\lambda$, diverges at the boundary of parameter space for $\lambda\rightarrow0$.}
\label{fig:S4}
\end{figure}

\subsection{Gauss distribution}
The KL divergence for continuous distributions is defined as
\begin{align}
    D_\mathrm{KL}(P||Q)=\int \mathrm d x\, P(x)\log \frac{P(x)}{Q(x)}\,.
\end{align}
The KL divergence between two Gaussian distributions with mean $\mu$ and variances $\sigma^2$ and $\sigma'^2$
\begin{align}
    P_{\mu,\,\sigma^2}(x)=\frac{1}{\sqrt{2\pi\sigma^2}}e^{-\frac{(x-\mu)^2}{2\sigma^2}}\,,\;P_{\mu,\,\sigma'^2}(x)=\frac{1}{\sqrt{2\pi\sigma'^2}}e^{-\frac{(x-\mu)^2}{2\sigma'^2}}\,,
\end{align}
is given by
\begin{align}
    D_\mathrm{KL}(P_{\mu,\,\sigma^2}||P_{\mu,\,\sigma'^2})=-\frac12\left[\log \sigma^2-\log\sigma'^2+1-\frac{\sigma^2}{\sigma'^2}\right]\,.
\end{align}
The next step is to Taylor expand the KL divergence around $\sigma$, by setting $\sigma'=\sigma+\delta\sigma$. For the Taylor expansion we find
\begin{align}
    D_\mathrm{KL}(P_{\mu,\,\sigma^2}||P_{\mu,\,\sigma'^2})&=\frac12\left[\partial_{\sigma'}^2 D_\mathrm{KL}(P_{\mu,\,\sigma^2}||P_{\mu,\,\sigma'^2})\right]_{\delta\sigma=0}\delta\sigma^2+O(\delta\sigma^3)=-\frac14\left[\frac{2}{\sigma'^2}-6\frac{\sigma^2}{\sigma'^4}\right]_{\delta\sigma=0}\delta\sigma^2+O(\delta\sigma^3)
\end{align}
From this we find  the FIM:
\begin{align}
    {\mathcal I}(\sigma)=\frac{1}{\sigma^2}
\end{align}
and  Jeffreys prior
\begin{align}\label{eq:Jeffreysprior_Gaussian}
    w(\sigma)=\frac{1}{\sigma}\,.
\end{align}
Jeffreys prior for the Gaussian is shown in Figure~\ref{fig:S5}. It diverges on the boundary of parameter space as $\sigma\rightarrow0$ and approaches zero for $\sigma\rightarrow\infty$.
\begin{figure}
\centering
\includegraphics[width=11.4cm]{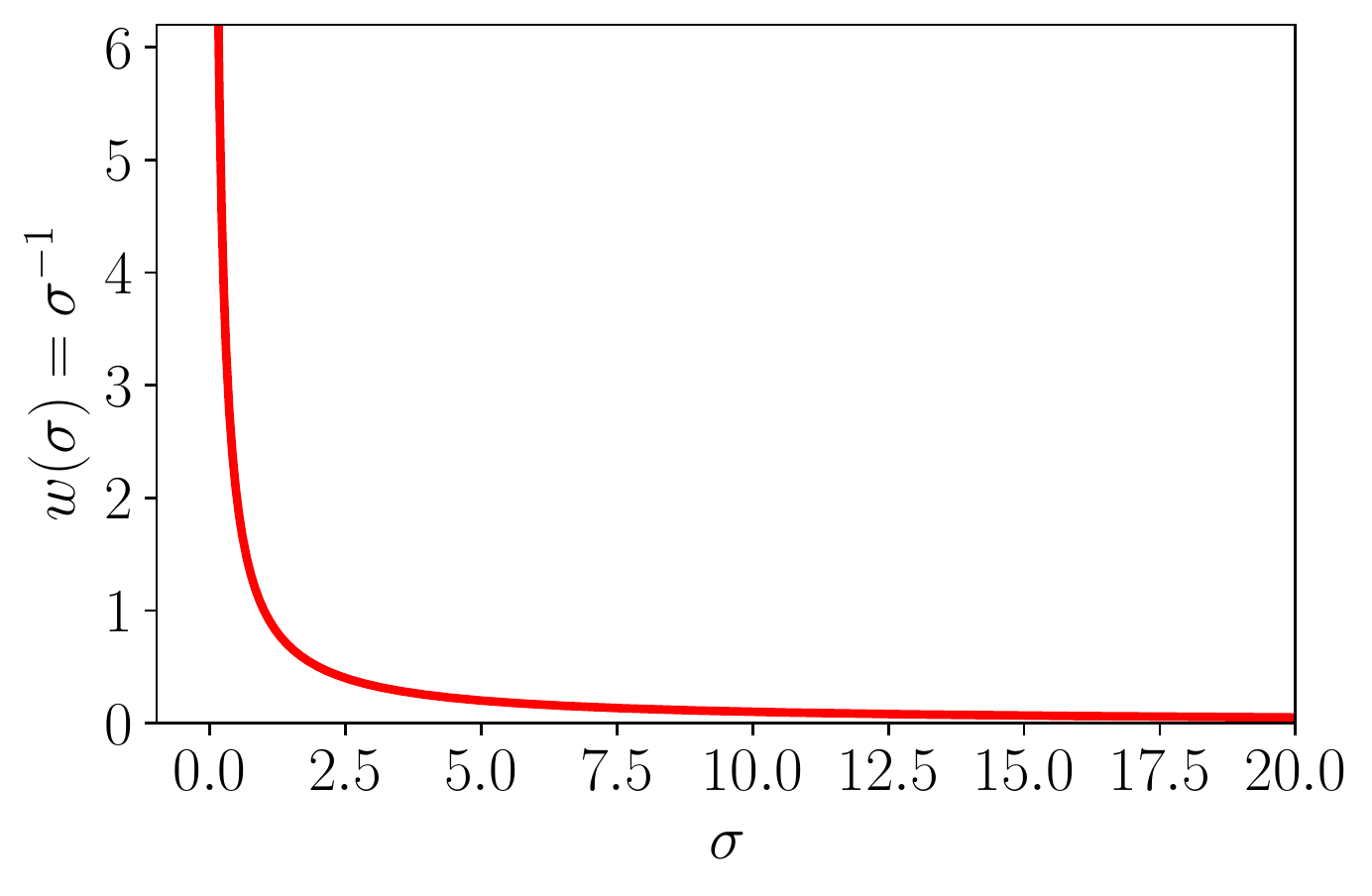}
\caption{Jeffreys prior for the Gaussian model with variance $\sigma^2$, diverges at the boundary of parameter space for $\sigma\rightarrow0$.}
\label{fig:S5}
\end{figure}

\section{Standard error and diverging KL divergence}
Consider the KL divergence for the Bernoulli model
\begin{align}
    D_\mathrm{KL}(P_{\theta^*}||P_{\theta})=H_{\theta^*}(\theta)-H_{\theta^*}(\theta^*)\,,
\end{align}
where $H_x(y)$ is the cross-entropy function
\begin{align}
    H_x(y)=-x\log y - (1-x)\log (1-y)\,.
\end{align}
For $\theta\rightarrow0,1$, the KL divergence diverges to plus infinity.
What is the behaviour of the KL divergence when we try to estimate the value of $\theta$ using data from many Bernoulli trials generated by the true model $\theta^*$? In particular, how does the KL divergence behave when $\theta^*$ lies very close to a boundary of parameter space? 

To answer this question we consider the empirically estimated parameter $\hat\theta$. This parameter is subject to statistical fluctuations due to finite data size. In general, the standard error for a parameter estimated from $N$ independent trials is given by
\begin{align}
    \sigma_N=\frac{\sigma}{\sqrt{N}}\,,
\end{align}
where $\sigma$ is the standard deviation of the distribution we are sampling from. 
Within the standard error the estimated parameter lies within the bounds
\begin{align}
    \hat\theta_\pm=\theta^*\pm \sigma_N\,.
\end{align}

For the Bernoulli model, the empirical estimate of the probability parameter is obtained as the mean number of events observed in $N$ independent Bernoulli trials and $\sigma$ is the standard deviation of the Bernoulli distribution
\begin{align}
    \sigma=\sqrt{\theta^*(1-\theta^*)}=\theta^*\sqrt{1/\theta^*-1}\,.
\end{align}
The statistically expected bounds are
\begin{align}
    \hat\theta_\pm=\theta^*\left(1-\sqrt{\frac{1/\theta^*-1}{N}}\right)\,.
\end{align}
Let us now focus on the boundary at $\theta=0$ and recall that the KL divergence diverges at this boundary. Therefore, one way to check whether we can expect the KL divergence to diverge for a given model $\theta^*$ and number of observations $N$, is to check whether the boundary $\theta=0$ lies within the standard error bound $\hat\theta_-$. From the last equation we see that the boundary is included if the condition
\begin{align}
    \sqrt{\frac{1/\theta^*-1}{N}}>1
\end{align}
is satisfied. Rearranging this inequality gives
\begin{align}
    \theta^*<\frac{1}{N+1}\,.
\end{align}
Thus true models which are within $1/(N+1)$ of the boundary, the boundary where the KL divergence diverges lies within the standard error and statistically we can expect the KL divergence to diverge. See Figure~\ref{fig:S1} for examples plots.\\

For the Poisson model, the empirical estimate of the parameter lambda is obtained as the mean number of events observed in $N$ independent Poisson trials and $\sigma$ is the standard deviation of the Poisson distribution
\begin{align}
    \sigma_N=\sqrt{\frac{\lambda^*}{N}}=\lambda^*\frac{1}{\sqrt{\lambda^* N}}\,.
\end{align}
The statistically expected bounds are
\begin{align}
    \hat\lambda_\pm=\lambda^* \pm\sigma_N=\lambda^*\left(1\pm\frac{1}{\sqrt{\lambda^* N}}\right)\,.
\end{align}
The boundary of parameter space lies at $\lambda=0$. Therefore we consider $\hat\lambda_-$ as the important bound and the condition for the boundary of parameter space to be included is
\begin{align}
    \frac{1}{\sqrt{\lambda^* N}}>1\,,
\end{align}
or
\begin{align}
    \lambda^*<\frac{1}{N}\,.
\end{align}
Therefore, for any given $N$ we can find a models $\lambda^*$ which lie close enough to the boundary according to the above condition, and for which we should expect the KL divergence to diverge.
\begin{figure}
\centering
\includegraphics[width=11.4cm]{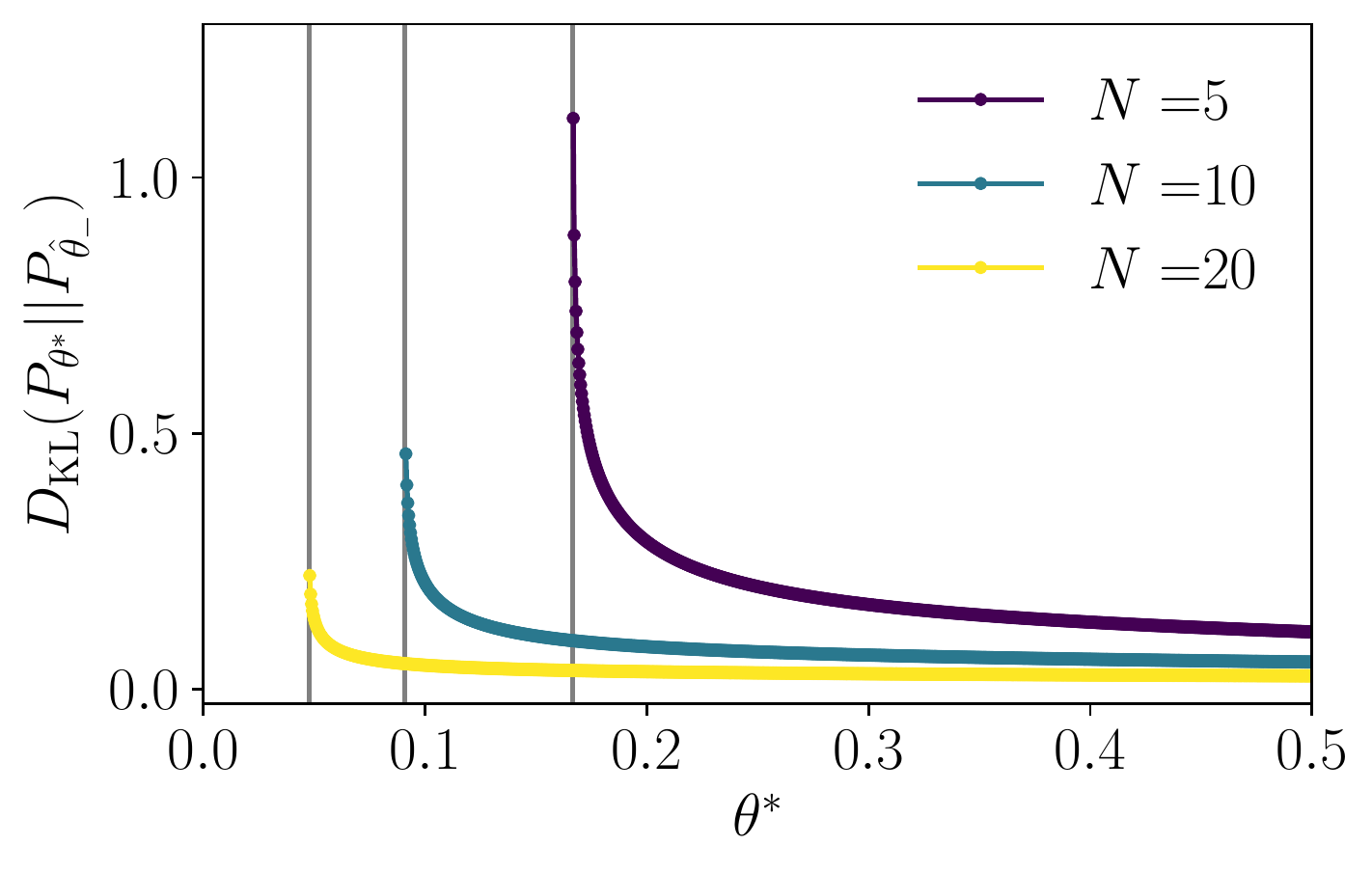}
\caption{Diverging behaviour of $D_\mathrm{KL}(P_{\theta^*}||P_{\hat\theta_-=\theta^*-\sigma_N})$ for three values of $N$. Vertical gray lines show the position of diverging behaviour at $1/(N+1)$.}
\label{fig:S1}
\end{figure}

\section{Optimality condition}

\subsection{The optimality condition for exponential families in mean parameterisation}
This derivation makes use of the fact that the Kullback-Leibler divergence between two distributions from the same exponential family in canonical form, can be expressed as the Bregman divergence (defined with respect to the dual of the log partition function, $F^*$) between the mean parameters $\boldsymbol\theta$:
\begin{align}
    D_\mathrm{KL}(Q_{\boldsymbol{\eta}^*}||Q_{\boldsymbol{\eta}})=B_{F^*}(\boldsymbol\theta^*:\boldsymbol\theta)\,.
\end{align}
Writing out the definition of the Bregman divergence, the right-hand side is of the form
\begin{align}\label{eq:Bregman}
    B_{F^*}(\boldsymbol{\theta}^*:\boldsymbol{\theta})=F^*(\boldsymbol{\theta}^*)-F^*(\boldsymbol{\theta})-[\boldsymbol{\theta}^*-\boldsymbol{\theta} ]^T\cdot\nabla_{\boldsymbol\theta} F^*(\boldsymbol{\theta})\,,
\end{align}
where
\begin{align}
    \boldsymbol\theta=\kappa \boldsymbol\theta_\mathrm{bias}+(1-\kappa)\hat{\boldsymbol\theta}_\mathrm{ML}\,.
\end{align}
To compute the condition for the optimal model, we take the derivative of the Bregman divergence with respect to the bias:
\begin{align}
    \frac{1}{\kappa}\nabla_{\theta_{\mathrm{bias}\,i}}B_{F^*}(\boldsymbol{\theta}^*:\boldsymbol{\theta})=\nabla_{\theta_i} B_{F^*}(\boldsymbol{\theta}^*:\boldsymbol{\theta})&=-\nabla_{\theta_i}F^*(\boldsymbol{\theta})-\boldsymbol{\theta}^{*T}\cdot\nabla_{\theta_i}\nabla_{\boldsymbol\theta} F^*(\boldsymbol{\theta} )+\nabla_{\theta_i}\{\boldsymbol{\theta}^T\cdot \nabla_{\boldsymbol\theta} F^*(\boldsymbol{\theta} )\}\\
    &=-\nabla_{\theta_i} F^*(\boldsymbol{\theta} )
    -\sum_{j=1}^d\theta^*_j \nabla_{\theta_i}\nabla_{\theta_j} F^*(\boldsymbol{\theta} )
    +\nabla_{\theta_i} F^*(\boldsymbol{\theta} )+\sum_{j=1}^d\theta_j \nabla_{\theta_i}\nabla_{\theta_j} F^*(\boldsymbol{\theta} )\\
    &=-\sum_{j=1}^d[\theta^*_j-\theta_j]\nabla_{\theta_i}\nabla_{\theta_j} F^*(\boldsymbol{\theta} )\,.
\end{align}
Next, we use the relation
\begin{align}
    \mathcal I_{ij} (\boldsymbol\theta) = \nabla_{\theta_i}\nabla_{\theta_j} F^*(\boldsymbol\theta)\,,
\end{align}
relating the Fisher information metric in mean parameterization to the dual function.
We derive this relation starting with \eqref{eq:DKL_Bregman}
\begin{align}
    D_\mathrm{KL}(Q_{\boldsymbol\eta^*(\boldsymbol\theta^*)}|| Q_{\boldsymbol\eta(\boldsymbol\theta)})=B_{F^*}({\boldsymbol\theta^*}:\boldsymbol\theta)\,,
\end{align}
replacing $\boldsymbol\theta=\boldsymbol\theta^*-\delta\boldsymbol\theta$, and Taylor expanding in $\delta\boldsymbol\theta$. Using the definition of the Bregman divergence~\eqref{eq:Bregman} and noting that $\nabla_{\delta\theta_i}=-\nabla_{\theta_i}$, the coefficients of the first three orders in $\delta\boldsymbol\theta$ are given by
\begin{align}
    B_{F^*}({\boldsymbol\theta^*}:\boldsymbol\theta^*-\delta\boldsymbol\theta)\big|_{\delta\boldsymbol\theta=0}&=0\,,\\
    \nabla_{\delta\theta_i}B_{F^*}({\boldsymbol\theta^*}:\boldsymbol\theta^*-\delta\boldsymbol\theta)\big|_{\delta\boldsymbol\theta=0}&=\left[\nabla_{\theta_i}F^*(\boldsymbol\theta)-\nabla_{\theta_i} F^*(\boldsymbol\theta)+\sum_j\delta\theta_j\nabla_{\theta_i}\nabla_{\theta)j}F^*(\boldsymbol\theta)\right]_{\delta\boldsymbol\theta=0}=0\,,\\
    \nabla_{\delta\theta_k}\nabla_{\delta\theta_i}B_{F^*}({\boldsymbol\theta^*}:\boldsymbol\theta^*-\delta\boldsymbol\theta)\big|_{\delta\boldsymbol\theta=0}&=\Big[-\nabla_{\delta\theta_k}\nabla_{\theta_i}F^*(\boldsymbol\theta)+\nabla_{\delta\theta_k}\nabla_{\theta_i} F^*(\boldsymbol\theta)
    +\delta_{kj}\nabla_{\theta_i}\nabla_{\theta_j}F^*(\boldsymbol\theta)
    \\&-\sum_j\delta\theta_j\nabla_{\delta\theta_k}\nabla_{\theta_i}\nabla_{\theta_j}F^*(\boldsymbol\theta)\Big]_{\delta\boldsymbol\theta=0}=\nabla_{\theta_i}\nabla_{\theta_k}F^*(\boldsymbol\theta^*)\,.
\end{align}{}
Using this relation, we rewrite the above condition as
\begin{align}
    \frac1\kappa\nabla_{\boldsymbol\theta_\mathrm{bias}}B_{F^*}(\boldsymbol{\theta}^*:\boldsymbol{\theta})=-\mathcal I (\boldsymbol\theta)\left[\boldsymbol{\theta}^*-\boldsymbol{\theta} \right]=\mathcal I(\boldsymbol\theta)\delta\boldsymbol
    \theta\,,
\end{align}
where
$-\delta\boldsymbol\theta\equiv\boldsymbol\theta^*-\boldsymbol\theta$. 
Thus, in mean parameterisation, the optimality constraint in Eq.~(6) of the main text, takes the simple form
\begin{align}
\frac{1}{\kappa}\nabla_{\boldsymbol\theta_\mathrm{bias}}\big< \mathcal L_{\boldsymbol\theta^*}(\boldsymbol{\theta})\big>_{\hat{\boldsymbol{\theta}}_\mathrm{ML}}=\big<\mathcal I(\boldsymbol\theta)\delta\boldsymbol\theta\big>_{\hat{\boldsymbol{\theta}}_\mathrm{ML}}=0\,.
\end{align}

\subsection{The optimality condition for the Bernoulli model in mean parameterization derived from Taylor expansion}
As an example, we evaluate the optimality condition
\begin{align}
    \nabla_{\boldsymbol\theta_\mathrm{bias}}\big<D_\mathrm{KL}(P_{\boldsymbol\theta^*}||P_{\boldsymbol\theta})\big>_{\hat{\boldsymbol\theta}_\mathrm{ML}}=0
\end{align}
for the Bernoulli model written in mean parameters. In this derivation we will use Taylor expansion of the KL divergence and find that all but a single term of the infinite series cancel to yield a simplified form of the condition in Eq.~(9) of main text.

We consider KL divergence between two Bernoulli distributions, denoted by $P_{\boldsymbol\theta^*}$ and $P_{\boldsymbol\theta}$, respectively:
\begin{align}
    D_\mathrm{KL}(P_{{\boldsymbol\theta}^*}||P_{{\boldsymbol\theta}})&=D_\mathrm{KL}(P_{{\boldsymbol\theta}+\delta{\boldsymbol\theta}}||P_{{\boldsymbol\theta}})=\sum_x P_{{\boldsymbol\theta}+\delta{\boldsymbol\theta}}\log\left(\frac{P_{{\boldsymbol\theta}+\delta{\boldsymbol\theta}}}{P_{{\boldsymbol\theta}}}\right)\,,
\end{align}
where ${\boldsymbol\theta}^*\equiv{\boldsymbol\theta}+\delta{\boldsymbol\theta}$ and expand the right-hand side in $\delta{\boldsymbol\theta}$. For the first few orders we find
\begin{align}
    D_\mathrm{KL}(P_{{\boldsymbol\theta}^*}||P_{{\boldsymbol\theta}})\Big|_{\delta{\boldsymbol\theta}=0}&=0\,,\\
    \partial D_\mathrm{KL}(P_{{\boldsymbol\theta}^*}||P_{{\boldsymbol\theta}})\Big|_{\delta{\boldsymbol\theta}=0}&=1\,,\\
    \partial^2 D_\mathrm{KL}(P_{{\boldsymbol\theta}^*}||P_{{\boldsymbol\theta}})\Big|_{\delta{\boldsymbol\theta}=0}&=1/{\boldsymbol\theta}\,,\\
    \partial^3 D_\mathrm{KL}(P_{{\boldsymbol\theta}^*}||P_{{\boldsymbol\theta}})\Big|_{\delta{\boldsymbol\theta}=0}&=-1/\boldsymbol\theta^2\,,\\
    \partial^4 D_\mathrm{KL}(P_{{\boldsymbol\theta}^*}||P_{{\boldsymbol\theta}})\Big|_{\delta{\boldsymbol\theta}=0}&=2/\boldsymbol\theta^3\,,\\
    \partial^5 D_\mathrm{KL}(P_{{\boldsymbol\theta}^*}||P_{{\boldsymbol\theta}})\Big|_{\delta{\boldsymbol\theta}=0}&=-6/\boldsymbol\theta^4\,,\\
    &\vdots\nonumber
\end{align}
where $\partial\equiv\partial_{\delta{\boldsymbol\theta}}$ and powers on the right-hand side act element-wise. The Taylor expansion is thus
\begin{align}
    D_\mathrm{KL}(P_{\boldsymbol\theta^*}||P_{\boldsymbol\theta})&=\sum_i\left\{\delta\theta_i+\frac{1}{2!}\frac{\delta\theta_i^2}{\theta_i}-\frac{1}{3!}\frac{\delta\theta_i^3}{\theta_i^2}+\frac{1}{4!}\frac{2\delta\theta_i^4}{\theta_i^3}-\frac{1}{5!}\frac{6\delta\theta_i^5}{\theta_i^4}+O(\delta_i^6)\right\}\\
    &=1+\sum_i\left\{\frac{1}{2!}\frac{\delta\theta_i^2}{\theta_i}-\frac{1}{3!}\frac{\delta\theta_i^3}{\theta_i^2}+\frac{1}{4!}\frac{2\delta\theta_i^4}{\theta_i^3}-\frac{1}{5!}\frac{6\delta\theta_i^5}{\theta_i^4}+O(\delta_i^6)\right\}
\end{align}
Next, we act with the derivative with respect to the bias:
\begin{align}
    \partial_{\theta_{\mathrm{bias}\,i}}D_\mathrm{KL}(P_{\boldsymbol\theta^*}||P_{\boldsymbol\theta})=(\partial\theta_i/\partial{\theta_{\mathrm{bias}\,i}})\partial_{\theta_i}D_\mathrm{KL}(P_{\boldsymbol\theta^*}||P_{\boldsymbol\theta})=\kappa\partial_{\theta_i}D_\mathrm{KL}(P_{\boldsymbol\theta^*}||P_{\boldsymbol\theta})\,.
\end{align}
Remembering that $\delta\boldsymbol\theta$ is a function of $\boldsymbol{\theta}$, $\partial\delta\theta/\partial\theta=-1$, we obtain:
\begin{align}
    \partial_{\theta_i}D_\mathrm{KL}(P_{\boldsymbol\theta^*}||P_{\boldsymbol\theta})&=-\frac{1}{\theta_i}\delta\theta_i+\left[-\frac{1}{2!}\frac{\delta\theta_i^2}{\theta_i^2}+\frac{3}{3!}\frac{\delta\theta_i^2}{\theta_i^2}\right]+\left[\frac{2}{3!}\frac{\delta\theta_i^3}{\theta_i^3}-\frac{4}{4!}\frac{2\delta\theta_i^3}{\theta_i^3}\right]+...\\
    &=-\frac{1}{\theta_i}\delta\theta_i\,,
\end{align}
where on in the first line on the right-hand side we have grouped terms such that expressions in the brackets vanish to show how terms of successive order in $\delta\theta$ cancel. Using $\theta_2=1-\theta_1$
, we finally see that in matrix notation the expression takes the form
\begin{align}
    \partial_{\boldsymbol\theta}D_\mathrm{KL}(P_{\boldsymbol\theta^*}||P_{\boldsymbol\theta})=-
    \begin{pmatrix}
1/\theta_1 & 0\\
0 & 1/(1-\theta_1)
\end{pmatrix}\delta\boldsymbol\theta
    =-\mathcal I(\boldsymbol\theta)\delta\boldsymbol\theta\,,
\end{align}
where we recognised the $2\times2$ matrix to be the Fisher information matrix. 
Thus, we obtain the optimality condition
\begin{align}
    -\big<\partial_{\boldsymbol\theta}D_\mathrm{KL}(P_{\boldsymbol\theta^*}||P_{\boldsymbol\theta})\big>_{\hat{\boldsymbol\theta}_\mathrm{ML}}
    =\big<\mathcal I(\boldsymbol\theta)\delta\boldsymbol\theta\big>_{\hat{\boldsymbol\theta}_\mathrm{ML}}=0\,.
\end{align}

\section{Optimal parameter calculation for Poisson distribution with unknown rate parameter}
\begin{figure*}[h]
\centering
\includegraphics[width=15.4cm]{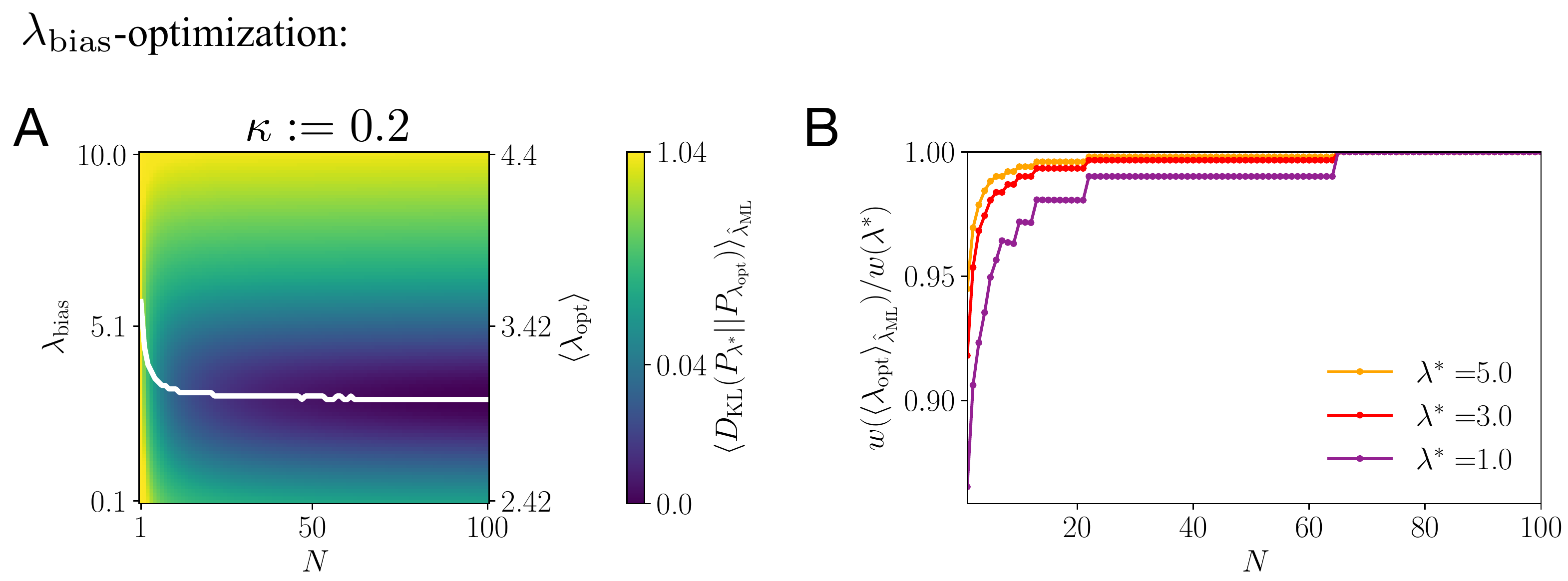}
\caption{Optimization of the expected loss for the Poisson distribution with respect to the prior shows a bias to low complexity models. The true model lies at $\lambda^*=3$.
(A) The bias $\lambda_\mathrm{bias}$ is optimised while $\kappa=0.2$ ($q=0.25$) is kept fixed. The plot shows the landscape of the expected loss as a function of the bias and the number of observations. We are using a non-linear colormap based on the cumulative distribution of the loss. (B) The plot shows for different true models $\lambda^*$ the ratio of the local complexity measure by Jeffreys prior of the expected optimal to the true model.
}
\label{fig:S2}
\end{figure*}
We consider two Poisson distributions with parameters $\lambda^*$ and $\lambda$:
\begin{align}
    P_{\lambda^*}(n)=\frac{\lambda^{*\,n}e^{-\lambda^*}}{n!}\,,\;
    P_{\lambda}(n)=\frac{\lambda^ne^{-\lambda}}{n!}\,.
\end{align}
The KL divergence between the two distributions is given by
\begin{align}
    D_\mathrm{KL}(P_{\lambda^*}||Q_\lambda)&=\sum_{n=0}^\infty\frac{\lambda^{*\,n}e^{-\lambda^*}}{n!}\log\left(\frac{\lambda^{*\,n}e^{-\lambda^*}}{n!}\frac{n!}{\lambda^ne^{-\lambda}}\right)=\lambda^*\log\frac{\lambda^*}{\lambda}+(\lambda-\lambda^*)\,.
\end{align}
We want to minimise the posterior mean loss
\begin{align}
    \big<\mathcal L_{\lambda^*}\big>_\mathrm{post}(\lambda)=\int \mathrm d \tilde\lambda P(\tilde\lambda|\mathbf n)D_\mathrm{KL}(P_{\tilde\lambda}||Q_\lambda)
\end{align}
with respect to $\lambda$. Taking the derivative with respect to $\lambda$ we obtain the condition
\begin{align}
\partial_\lambda  \big<\mathcal L_{\lambda^*}\big>_\mathrm{post}=\partial_{\lambda} \int \mathrm d \tilde\lambda P(\tilde\lambda|\mathbf n)D_\mathrm{KL}(P_{\tilde\lambda}||Q_\lambda)=\int \mathrm d \tilde\lambda P(\tilde\lambda|\mathbf n)\left(-\tilde\lambda/\lambda+1\right)=0\,.
\end{align}
The condition is satisfied by
\begin{align}
    \lambda =\int \mathrm d \tilde\lambda P(\tilde\lambda|\mathbf n) \tilde\lambda\,.
\end{align}
The conjugate prior for the mean number of counts $\lambda$ is the Gamma distribution and the posterior takes the form
\begin{align}
    P(\lambda|\mathbf n)=\mathrm{Gamma}(\alpha',\beta')=\frac{{\beta'}^{\alpha'}}{\Gamma(\alpha')}\lambda^{\alpha'-1}e^{-\beta'\lambda}
\end{align}
with
\begin{align}
    \alpha'&=\alpha+N\bar X = \alpha + N \frac{1}{N}\sum_{i=1}^N x_i\,,\\
    \beta'&=\beta + N\,,
\end{align}
and the hyperparameters $\alpha,\beta>0$. $\bar X$ is an empirical estimate of the rate $\lambda$ from $N$ observations. The posterior mean which gives the optimal parameter choice is given by
\begin{align}
    \lambda(\hat\lambda_\mathrm{ML};\alpha,\beta)=\frac{\alpha'}{\beta'}=\frac{\alpha+N\hat\lambda_\mathrm{ML}}{\beta+N}=\frac{\frac\beta N\frac\alpha \beta+\hat\lambda_\mathrm{ML}}{\frac\beta N + 1}=\frac{q\lambda_\mathrm{bias}+\hat\lambda_\mathrm{ML}}{q+1}
\end{align}
and we define as before
\begin{align}
    N_0\equiv\beta\,,\quad \lambda_\mathrm{bias}\equiv\frac{\alpha}{\beta}\,,\quad\text{and }\quad q\equiv\frac{\beta}{N}\,.
\end{align}
The hyperparameter $\beta$ can thus be interpreted as virtual observations and $\alpha$ as the sum of counts across $\beta$ observations. If we do not trust in the empirical estimate, then we should take $\alpha$ large compared to $\beta$. In this limit, the optimal $\lambda$ values becomes large and lies further away from the boundary of parameter space situated at $\lambda=0$.

Up to a scaling by $1/N$, the maximum-likelihood value of the average count $\hat\lambda_\mathrm{ML}$ computed from $N$ observations, is distributed according to the true Poisson distribution with the rate rescaled by $N$. This can be seen from the characteristic function of the Poisson distribution given by $\exp(\lambda(e^{it}-1))$ and the fact that the characteristic function of the sum of $N$ i.i.d. random variables is given by the $N$'th power of the characteristic function of a single random variable. The empirically averaged expected loss is then given by
\begin{align}
    \big<\mathcal L_{\lambda^*}(\lambda)\big>_{\hat\lambda_\mathrm{ML}}=\sum_{n=0}^\infty\, \mathrm{Poi}(n;N\lambda^*)D_\mathrm{KL}(P_{\lambda^*}||Q_{{\lambda}(n/N;\alpha,\beta)})\,.
\end{align}
We minimise this loss with respect to the hyperparameters, for instance by keeping $\beta$ fixed an minimising with respect to $\alpha$. In Figure~\ref{fig:S2} we show the optimal choice of the hyperparameter $\alpha$ as a function of the data size, confirming that for small $N$, $\alpha$ is large, while for increasing data size $\alpha\rightarrow0$.

\section{Optimal parameter calculation for Gaussian with known mean and unknown variance}
Consider two Gaussian distributions with the same known mean $\mu$ and unknown variances ${\sigma^*}^2$ and $\sigma^2$
\begin{align}
    P_{\sigma^{*2}}(x)=\frac{1}{\sqrt{2\pi{\sigma^*}^2}}e^{-\frac{(x-\mu)^2}{2{\sigma^*}^2}}\,,\;P_{\sigma^2}(x)=\frac{1}{\sqrt{2\pi\sigma^2}}e^{-\frac{(x-\mu)^2}{2\sigma^2}}\,.
\end{align}
The KL divergence two between these distributions is given by
\begin{align}
    D_\mathrm{KL}(P_{\sigma^{*2}}||P_{\sigma^2})=-\frac12\left[\log {\sigma^*}^2-\log\sigma^2+1-\frac{{\sigma^*}^2}{\sigma^2}\right]\,.
\end{align}
We want to minimise the posterior mean loss
\begin{align}
     \big<\mathcal L_{\sigma^*}\big>_\mathrm{post}(\sigma)=\int \mathrm d {\tilde\sigma}^2 P({\tilde\sigma}^2|\mathbf n)D_\mathrm{KL}(P_{\tilde\sigma^{2}}||P_{\sigma^2})
\end{align}
with respect to $\sigma^2$, where $P({\tilde\sigma}^2|\mathbf n)$ is the posterior.
Taking the derivative with respect to $\sigma$ we obtain the condition
\begin{align}
    \partial_{\sigma}\int \mathrm d {\tilde\sigma}^2 P({\tilde\sigma}^2|\mathbf n)D_\mathrm{KL}(P_{\tilde\sigma^2}||P_{\sigma^2})=\int \mathrm d {\tilde\sigma}^2 P({\tilde\sigma}^2|\mathbf n)\left(\frac{1}{\sigma}-\frac{{\tilde\sigma}^2}{\sigma^3}\right)=0
\end{align}
satisfied by
\begin{align}
    \sigma^2=\int \mathrm d {\tilde\sigma}^2 {\tilde\sigma}^2 P({\tilde\sigma}^2|\mathbf n)\,.
\end{align}
\begin{figure*}[h]
\centering
\includegraphics[width=15.4cm]{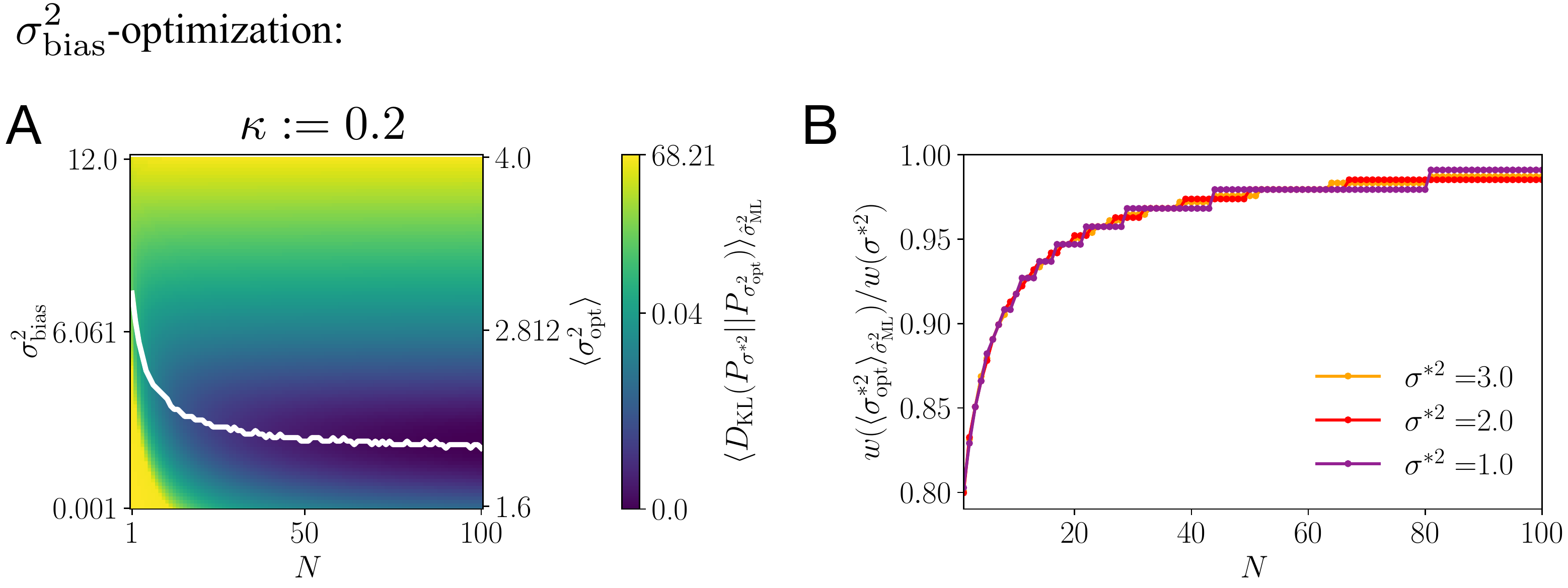}
\caption{Optimization of the expected loss for the Gaussian distribution with respect to the prior shows a bias to low complexity models. The true model lies at $\sigma^{*2}=2$ and we set $\mu=0$ for the known mean value.
(A) The bias $\sigma_\mathrm{bias}^2$ is optimised while $\kappa=0.2$ ($q=0.25$) is kept fixed. The plot shows the landscape of the expected loss as a function of the bias and the number of observations. We are using a non-linear colormap based on the cumulative distribution of the loss. (B) The plot shows for different true models $\sigma^{*2}$ the ratio of the local complexity measure by Jeffreys prior of the expected optimal to the true model.
}
\label{fig:S3}
\end{figure*}
The conjugate prior for the variance is the scaled inverse-chi-squared distribution $\chi^{-2}$. In \cite{KevinMurphy_Gaussian}, the posterior distribution for the variance, is given as
\begin{align}
    P(\sigma^2|\nu_N,\sigma_N^2)=\chi^{-2}(\sigma^2|\nu_N,\sigma_N^2)=\frac{1}{\Gamma(\nu_N/2)}\left(\frac{\nu_N{\sigma}^2_N}{2}\right)^{\nu_N/2}x^{-\nu_N/2-1}e^{-\frac{\nu_N\sigma_N^2}{2x}}\,,
\end{align}
where
\begin{align}
    \nu_N&=\nu+N\,,\quad\text{with}\quad \nu>2\,,\\
    \sigma_N^2&=\frac{1}{\nu_N}\left(\nu \sigma_0^2+\sum_{i=1}^N(x_i-\mu)^2\right)\,.
\end{align}
The loss-minimizing value for $\sigma^2$ is given by the Bayesian mean which comes out as
\begin{align}
    \sigma^2(\hat\sigma_\mathrm{ML}^2;\nu,\sigma_0^2)=\int \mathrm d {\tilde\sigma}^2 {\tilde\sigma}^2 \chi^{-2}(\tilde\sigma^2|\nu_N,\sigma_N^2)&=\frac{\nu_N}{\nu_N-2}\sigma_N^2=\frac{\nu\sigma_0^2+\sum_i(x_i-\mu)^2}{\nu-2+N}=\frac{\frac{\nu-2}{N}\frac{\nu\sigma_0^2}{\nu-2}+\frac{1}{N}\sum_i(x_i-\mu)^2}{\frac{\nu-2}{N}+1}
\end{align}
and we define
\begin{align}
    N_0&\equiv \nu-2\,,\\
    \sigma_\mathrm{bias}^2&\equiv\frac{\nu\sigma_0^2}{\nu-2}\,,\\
    \hat\sigma_\mathrm{ML}^2&\equiv\frac1N\sum_i(x_i-\mu)^2\,.
\end{align}

Up to a scaling of $\sigma^{*2}/N$, the empirical variance is distributed according to the chi-squared distribution of order $N$ (recall, the variance is given by the sum of $N$ i.i.d. squared Gaussian random variables divided by $N$) given by
\begin{align}
    \chi^2_N(x)=\frac {1}{2^{N/2}\Gamma (N/2)}\;x^{N/2-1}e^{-x/2}\,.
\end{align}
Finally, we optimise the expected loss
\begin{align}
    \big<\mathcal L_{\sigma^*}(\sigma)\big>_{\hat\sigma_\mathrm{ML}}=\int \mathrm d x\, \chi^2_N(x)D_\mathrm{KL}(P_{{\sigma^*}^2}||P_{{\sigma}^2(\sigma^{*2}x/N;\nu,\sigma_0^2)})
\end{align}
with respect to the bias term, while $\kappa$ is fixed. In this equation we are using the continuous version of the Kullback-Leibler divergence. By Eq.~(18) of the main text, $\kappa$ fixes $q$ which in turn provides a value for $N_0$ given $N$:
\begin{align}
    q=\frac{N_0}{N}=\frac{\kappa}{1-\kappa}\,,\quad N_0=Nq=\nu-2\,,
\end{align}
where we have also used the definition of $N_0$ in terms of the hyperparameter $\nu$. For the hyperparameters $\nu$ and the bias we can thus write:
\begin{align}
    \nu&=N_0+2\,,\\
    \sigma_\mathrm{bias}^2&=\frac{N_0+2}{N_0}\sigma_0^2\,.
\end{align}
Minimising the expected loss given $N$, yields the optimal choice for $\sigma_\mathrm{bias}^2$. In Figure~\ref{fig:S3} we show the optimal bias value as a function of the data size.


\section{Statistical fluctuations of different learning strategies}
\begin{figure*}
\centering
\includegraphics[width=15.4cm]{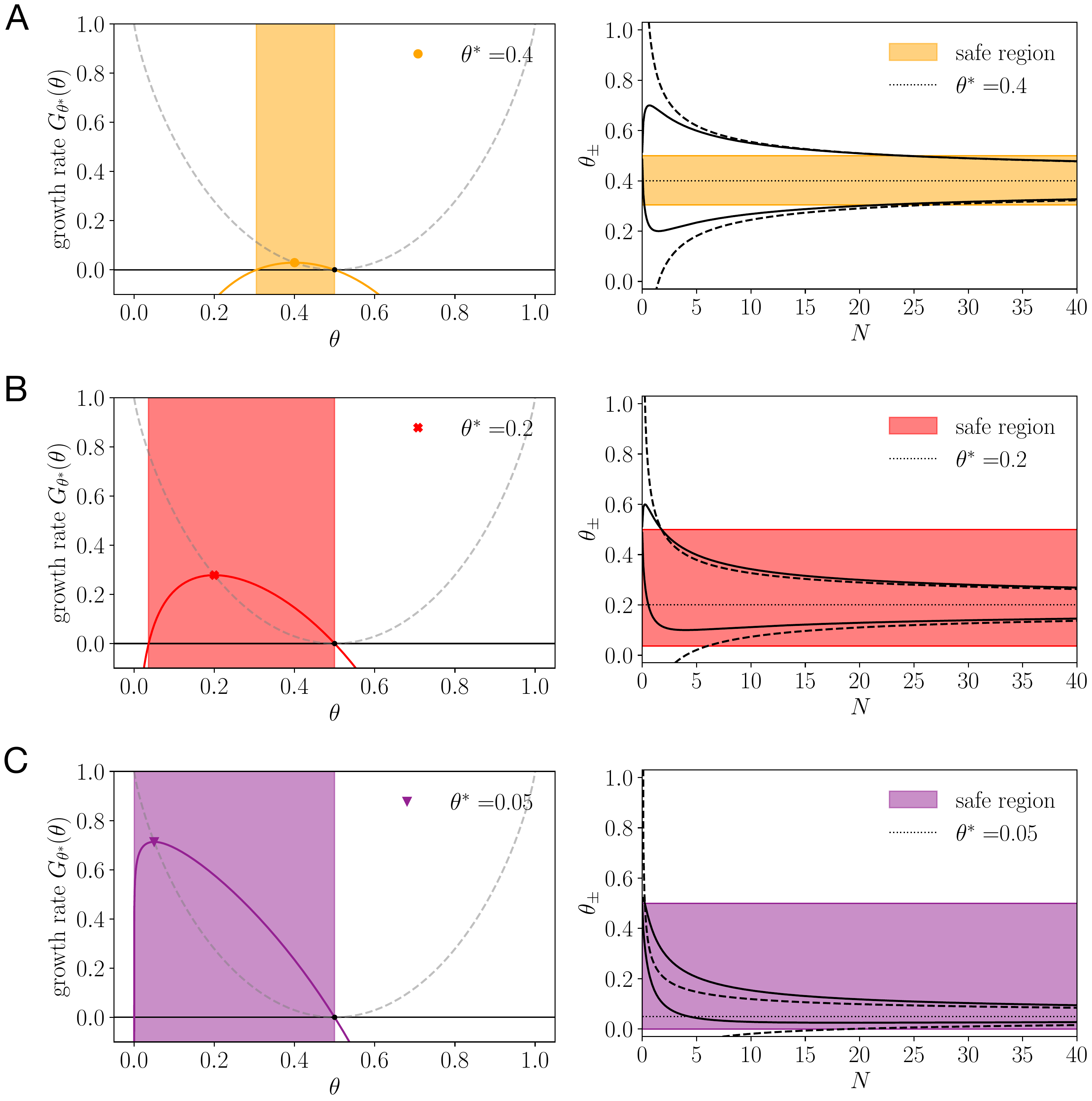}
\caption{The prior choice in Bayesian inference controls the safety of the learning strategy. Left column: long-term growth rate as a function of the adopted model $\phi$ for three different values of the true model $\theta^*=0.4$, $0.2$ and $0.05$. Shaded boxes indicate the regions of non-negative growth rate. Right column: Magnitude of fluctuations of the conditional optimal model as a function of the number of observations $N$ for two different learning strategies: maximum-likelihood estimation ($\alpha=0$; dashed line) and Jeffreys prior ($\alpha=1/2$; solid line).}
\label{fig:S7}
\end{figure*}
In Figure~\ref{fig:S7}, we show the magnitude of statistical fluctuations of the inferred model in relation to the region of non-negative long-term growth rates. Subfigures A, B and C show different true models $\theta^*=0.4$, $0.2$ and $0.05$, respectively. In the left column of subfigures we show the growth rates $G_{\theta^*}(\theta)$ as a function of the model choice $\theta$. Regions where the growth rate is non-negative are indicated by color shaded boxes. In the right column of subfigures we show the statistical fluctuations of the inferred model as function of $N$ for two different learning strategies. The first learning strategy is maximum-likelihood estimation $\alpha=0$ and the second strategy uses $\alpha=1/2$. The later strategy ($\alpha=1/2$) is safer, because the magnitude of fluctuations is shrunk and more contained in the interval of non-negative growth rates in comparison to the maximum-likelihood strategy.

\end{widetext}

\newpage

\bibliography{apssamp}

\providecommand{\noopsort}[1]{}\providecommand{\singleletter}[1]{#1}%
\begin{thebibliography}{56}%
\makeatletter
\providecommand \@ifxundefined [1]{%
 \@ifx{#1\undefined}
}%
\providecommand \@ifnum [1]{%
 \ifnum #1\expandafter \@firstoftwo
 \else \expandafter \@secondoftwo
 \fi
}%
\providecommand \@ifx [1]{%
 \ifx #1\expandafter \@firstoftwo
 \else \expandafter \@secondoftwo
 \fi
}%
\providecommand \natexlab [1]{#1}%
\providecommand \enquote  [1]{``#1''}%
\providecommand \bibnamefont  [1]{#1}%
\providecommand \bibfnamefont [1]{#1}%
\providecommand \citenamefont [1]{#1}%
\providecommand \href@noop [0]{\@secondoftwo}%
\providecommand \href [0]{\begingroup \@sanitize@url \@href}%
\providecommand \@href[1]{\@@startlink{#1}\@@href}%
\providecommand \@@href[1]{\endgroup#1\@@endlink}%
\providecommand \@sanitize@url [0]{\catcode `\\12\catcode `\$12\catcode
  `\&12\catcode `\#12\catcode `\^12\catcode `\_12\catcode `\%12\relax}%
\providecommand \@@startlink[1]{}%
\providecommand \@@endlink[0]{}%
\providecommand \url  [0]{\begingroup\@sanitize@url \@url }%
\providecommand \@url [1]{\endgroup\@href {#1}{\urlprefix }}%
\providecommand \urlprefix  [0]{URL }%
\providecommand \Eprint [0]{\href }%
\providecommand \doibase [0]{https://doi.org/}%
\providecommand \selectlanguage [0]{\@gobble}%
\providecommand \bibinfo  [0]{\@secondoftwo}%
\providecommand \bibfield  [0]{\@secondoftwo}%
\providecommand \translation [1]{[#1]}%
\providecommand \BibitemOpen [0]{}%
\providecommand \bibitemStop [0]{}%
\providecommand \bibitemNoStop [0]{.\EOS\space}%
\providecommand \EOS [0]{\spacefactor3000\relax}%
\providecommand \BibitemShut  [1]{\csname bibitem#1\endcsname}%
\let\auto@bib@innerbib\@empty
\bibitem [{\citenamefont {Burnet}(1976)}]{Burnet}%
  \BibitemOpen
  \bibfield  {author} {\bibinfo {author} {\bibfnamefont {F.~M.}\ \bibnamefont
  {Burnet}},\ }\bibfield  {title} {\bibinfo {title} {A modification of
  {J}erne's theory of antibody production using the concept of clonal
  selection},\ }\href
  {https://doi.org/https://doi.org/10.3322/canjclin.26.2.119} {\bibfield
  {journal} {\bibinfo  {journal} {CA: A Cancer Journal for Clinicians}\
  }\textbf {\bibinfo {volume} {26}},\ \bibinfo {pages} {119} (\bibinfo {year}
  {1976})}\BibitemShut {NoStop}%
\bibitem [{\citenamefont {Kussell}\ and\ \citenamefont
  {Leibler}(2005)}]{Kussell2075}%
  \BibitemOpen
  \bibfield  {author} {\bibinfo {author} {\bibfnamefont {E.}~\bibnamefont
  {Kussell}}\ and\ \bibinfo {author} {\bibfnamefont {S.}~\bibnamefont
  {Leibler}},\ }\bibfield  {title} {\bibinfo {title} {Phenotypic diversity,
  population growth, and information in fluctuating environments},\ }\href
  {https://doi.org/10.1126/science.1114383} {\bibfield  {journal} {\bibinfo
  {journal} {Science}\ }\textbf {\bibinfo {volume} {309}},\ \bibinfo {pages}
  {2075} (\bibinfo {year} {2005})}\BibitemShut {NoStop}%
\bibitem [{\citenamefont {Xue}\ and\ \citenamefont {Leibler}(2018)}]{Xue12745}%
  \BibitemOpen
  \bibfield  {author} {\bibinfo {author} {\bibfnamefont {B.}~\bibnamefont
  {Xue}}\ and\ \bibinfo {author} {\bibfnamefont {S.}~\bibnamefont {Leibler}},\
  }\bibfield  {title} {\bibinfo {title} {Benefits of phenotypic plasticity for
  population growth in varying environments},\ }\href
  {https://doi.org/10.1073/pnas.1813447115} {\bibfield  {journal} {\bibinfo
  {journal} {Proceedings of the National Academy of Sciences}\ }\textbf
  {\bibinfo {volume} {115}},\ \bibinfo {pages} {12745} (\bibinfo {year}
  {2018})}\BibitemShut {NoStop}%
\bibitem [{\citenamefont {Hopfield}(1974)}]{proof_reading}%
  \BibitemOpen
  \bibfield  {author} {\bibinfo {author} {\bibfnamefont {J.~J.}\ \bibnamefont
  {Hopfield}},\ }\bibfield  {title} {\bibinfo {title} {Kinetic proofreading: A
  new mechanism for reducing errors in biosynthetic processes requiring high
  specificity},\ }\href {https://doi.org/10.1073/pnas.71.10.4135} {\bibfield
  {journal} {\bibinfo  {journal} {Proceedings of the National Academy of
  Sciences}\ }\textbf {\bibinfo {volume} {71}},\ \bibinfo {pages} {4135}
  (\bibinfo {year} {1974})}\BibitemShut {NoStop}%
\bibitem [{\citenamefont {Martincorena}\ \emph {et~al.}(2012)\citenamefont
  {Martincorena}, \citenamefont {Seshasayee},\ and\ \citenamefont
  {Luscombe}}]{martincorena2012evidence}%
  \BibitemOpen
  \bibfield  {author} {\bibinfo {author} {\bibfnamefont {I.}~\bibnamefont
  {Martincorena}}, \bibinfo {author} {\bibfnamefont {A.~S.}\ \bibnamefont
  {Seshasayee}},\ and\ \bibinfo {author} {\bibfnamefont {N.~M.}\ \bibnamefont
  {Luscombe}},\ }\bibfield  {title} {\bibinfo {title} {Evidence of non-random
  mutation rates suggests an evolutionary risk management strategy},\
  }\href@noop {} {\bibfield  {journal} {\bibinfo  {journal} {Nature}\ }\textbf
  {\bibinfo {volume} {485}},\ \bibinfo {pages} {95} (\bibinfo {year}
  {2012})}\BibitemShut {NoStop}%
\bibitem [{\citenamefont {Harder}\ and\ \citenamefont {Real}(1987)}]{bees}%
  \BibitemOpen
  \bibfield  {author} {\bibinfo {author} {\bibfnamefont {L.~D.}\ \bibnamefont
  {Harder}}\ and\ \bibinfo {author} {\bibfnamefont {L.~A.}\ \bibnamefont
  {Real}},\ }\bibfield  {title} {\bibinfo {title} {Why are bumble bees risk
  averse?},\ }\href {https://doi.org/https://doi.org/10.2307/1938384}
  {\bibfield  {journal} {\bibinfo  {journal} {Ecology}\ }\textbf {\bibinfo
  {volume} {68}},\ \bibinfo {pages} {1104} (\bibinfo {year}
  {1987})}\BibitemShut {NoStop}%
\bibitem [{\citenamefont {Bhat}\ \emph {et~al.}(2020)\citenamefont {Bhat},
  \citenamefont {Kempes},\ and\ \citenamefont {Yeakel}}]{life_histories}%
  \BibitemOpen
  \bibfield  {author} {\bibinfo {author} {\bibfnamefont {U.}~\bibnamefont
  {Bhat}}, \bibinfo {author} {\bibfnamefont {C.~P.}\ \bibnamefont {Kempes}},\
  and\ \bibinfo {author} {\bibfnamefont {J.~D.}\ \bibnamefont {Yeakel}},\
  }\bibfield  {title} {\bibinfo {title} {Scaling the risk landscape drives
  optimal life-history strategies and the evolution of grazing},\ }\href
  {https://doi.org/10.1073/pnas.1907998117} {\bibfield  {journal} {\bibinfo
  {journal} {Proceedings of the National Academy of Sciences}\ }\textbf
  {\bibinfo {volume} {117}},\ \bibinfo {pages} {1580} (\bibinfo {year}
  {2020})}\BibitemShut {NoStop}%
\bibitem [{\citenamefont {Sutton}\ and\ \citenamefont
  {Barto}(2018)}]{sutton2018reinforcement}%
  \BibitemOpen
  \bibfield  {author} {\bibinfo {author} {\bibfnamefont {R.~S.}\ \bibnamefont
  {Sutton}}\ and\ \bibinfo {author} {\bibfnamefont {A.~G.}\ \bibnamefont
  {Barto}},\ }\href@noop {} {\emph {\bibinfo {title} {Reinforcement learning:
  An introduction}}}\ (\bibinfo  {publisher} {MIT press},\ \bibinfo {year}
  {2018})\BibitemShut {NoStop}%
\bibitem [{\citenamefont {Vergassola}\ \emph {et~al.}(2007)\citenamefont
  {Vergassola}, \citenamefont {Villermaux},\ and\ \citenamefont
  {Shraiman}}]{vergassola2007infotaxis}%
  \BibitemOpen
  \bibfield  {author} {\bibinfo {author} {\bibfnamefont {M.}~\bibnamefont
  {Vergassola}}, \bibinfo {author} {\bibfnamefont {E.}~\bibnamefont
  {Villermaux}},\ and\ \bibinfo {author} {\bibfnamefont {B.~I.}\ \bibnamefont
  {Shraiman}},\ }\bibfield  {title} {\bibinfo {title} {‘infotaxis’ as a
  strategy for searching without gradients},\ }\href@noop {} {\bibfield
  {journal} {\bibinfo  {journal} {Nature}\ }\textbf {\bibinfo {volume} {445}},\
  \bibinfo {pages} {406} (\bibinfo {year} {2007})}\BibitemShut {NoStop}%
\bibitem [{\citenamefont {Cohen}(1966)}]{Cohen_1966}%
  \BibitemOpen
  \bibfield  {author} {\bibinfo {author} {\bibfnamefont {D.}~\bibnamefont
  {Cohen}},\ }\bibfield  {title} {\bibinfo {title} {Optimizing reproduction in
  a randomly varying environment},\ }\href
  {https://doi.org/https://doi.org/10.1016/0022-5193(66)90188-3} {\bibfield
  {journal} {\bibinfo  {journal} {Journal of Theoretical Biology}\ }\textbf
  {\bibinfo {volume} {12}},\ \bibinfo {pages} {119} (\bibinfo {year}
  {1966})}\BibitemShut {NoStop}%
\bibitem [{\citenamefont {Rubenstein}(2011)}]{cooperative_breeding}%
  \BibitemOpen
  \bibfield  {author} {\bibinfo {author} {\bibfnamefont {D.~R.}\ \bibnamefont
  {Rubenstein}},\ }\bibfield  {title} {\bibinfo {title} {Spatiotemporal
  environmental variation, risk aversion, and the evolution of cooperative
  breeding as a bet-hedging strategy},\ }\href
  {https://doi.org/10.1073/pnas.1100303108} {\bibfield  {journal} {\bibinfo
  {journal} {Proceedings of the National Academy of Sciences}\ }\textbf
  {\bibinfo {volume} {108}},\ \bibinfo {pages} {10816} (\bibinfo {year}
  {2011})}\BibitemShut {NoStop}%
\bibitem [{\citenamefont {Levins}(1962)}]{Levins_1962}%
  \BibitemOpen
  \bibfield  {author} {\bibinfo {author} {\bibfnamefont {R.}~\bibnamefont
  {Levins}},\ }\bibfield  {title} {\bibinfo {title} {Theory of fitness in a
  heterogeneous environment. i. the fitness set and adaptive function},\ }\href
  {https://doi.org/10.1086/282245} {\bibfield  {journal} {\bibinfo  {journal}
  {The American Naturalist}\ }\textbf {\bibinfo {volume} {96}},\ \bibinfo
  {pages} {361} (\bibinfo {year} {1962})}\BibitemShut {NoStop}%
\bibitem [{\citenamefont {Nemenman}(2012)}]{nemenman2012information}%
  \BibitemOpen
  \bibfield  {author} {\bibinfo {author} {\bibfnamefont {I.}~\bibnamefont
  {Nemenman}},\ }\bibfield  {title} {\bibinfo {title} {Information theory and
  adaptation},\ }\href@noop {} {\bibfield  {journal} {\bibinfo  {journal}
  {Quantitative biology: from molecular to cellular systems}\ }\textbf
  {\bibinfo {volume} {4}},\ \bibinfo {pages} {73} (\bibinfo {year}
  {2012})}\BibitemShut {NoStop}%
\bibitem [{\citenamefont {Perkins}\ and\ \citenamefont
  {Swain}(2009)}]{perkins2009strategies}%
  \BibitemOpen
  \bibfield  {author} {\bibinfo {author} {\bibfnamefont {T.~J.}\ \bibnamefont
  {Perkins}}\ and\ \bibinfo {author} {\bibfnamefont {P.~S.}\ \bibnamefont
  {Swain}},\ }\bibfield  {title} {\bibinfo {title} {Strategies for cellular
  decision-making},\ }\href@noop {} {\bibfield  {journal} {\bibinfo  {journal}
  {Molecular systems biology}\ }\textbf {\bibinfo {volume} {5}},\ \bibinfo
  {pages} {326} (\bibinfo {year} {2009})}\BibitemShut {NoStop}%
\bibitem [{\citenamefont {{Laplace, marquis de}}(1814)}]{Laplace_1814}%
  \BibitemOpen
  \bibfield  {author} {\bibinfo {author} {\bibfnamefont {P.~S.}\ \bibnamefont
  {{Laplace, marquis de}}},\ }\href@noop {} {\emph {\bibinfo {title} {{E}ssai
  philosophique sur les probabilités}}}\ (\bibinfo  {publisher} {Courcier},\
  \bibinfo {year} {1814})\BibitemShut {NoStop}%
\bibitem [{\citenamefont {Keynes}(1921)}]{Keynes_1921}%
  \BibitemOpen
  \bibfield  {author} {\bibinfo {author} {\bibfnamefont {J.~M.}\ \bibnamefont
  {Keynes}},\ }\href@noop {} {\emph {\bibinfo {title} {A Treatise on
  Probability}}}\ (\bibinfo  {publisher} {Dover Publications},\ \bibinfo {year}
  {1921})\BibitemShut {NoStop}%
\bibitem [{\citenamefont {Kullback}(1997)}]{kullback1997information}%
  \BibitemOpen
  \bibfield  {author} {\bibinfo {author} {\bibfnamefont {S.}~\bibnamefont
  {Kullback}},\ }\href@noop {} {\emph {\bibinfo {title} {Information theory and
  statistics}}}\ (\bibinfo  {publisher} {Courier Corporation},\ \bibinfo {year}
  {1997})\BibitemShut {NoStop}%
\bibitem [{\citenamefont {Jaynes}(2003)}]{Jaynes_2003}%
  \BibitemOpen
  \bibfield  {author} {\bibinfo {author} {\bibfnamefont {E.~T.}\ \bibnamefont
  {Jaynes}},\ }\href {https://doi.org/10.1017/CBO9780511790423} {\emph
  {\bibinfo {title} {Probability Theory: The Logic of Science}}},\ edited by\
  \bibinfo {editor} {\bibfnamefont {G.~L.}\ \bibnamefont {Bretthorst}}\
  (\bibinfo  {publisher} {Cambridge University Press},\ \bibinfo {year}
  {2003})\BibitemShut {NoStop}%
\bibitem [{\citenamefont {Jaynes}(1957{\natexlab{a}})}]{MaxEnt_I}%
  \BibitemOpen
  \bibfield  {author} {\bibinfo {author} {\bibfnamefont {E.~T.}\ \bibnamefont
  {Jaynes}},\ }\bibfield  {title} {\bibinfo {title} {Information theory and
  statistical mechanics},\ }\href {https://doi.org/10.1103/PhysRev.106.620}
  {\bibfield  {journal} {\bibinfo  {journal} {Phys. Rev.}\ }\textbf {\bibinfo
  {volume} {106}},\ \bibinfo {pages} {620} (\bibinfo {year}
  {1957}{\natexlab{a}})}\BibitemShut {NoStop}%
\bibitem [{\citenamefont {Jaynes}(1957{\natexlab{b}})}]{MaxEnt_II}%
  \BibitemOpen
  \bibfield  {author} {\bibinfo {author} {\bibfnamefont {E.~T.}\ \bibnamefont
  {Jaynes}},\ }\bibfield  {title} {\bibinfo {title} {Information theory and
  statistical mechanics. ii},\ }\href {https://doi.org/10.1103/PhysRev.108.171}
  {\bibfield  {journal} {\bibinfo  {journal} {Phys. Rev.}\ }\textbf {\bibinfo
  {volume} {108}},\ \bibinfo {pages} {171} (\bibinfo {year}
  {1957}{\natexlab{b}})}\BibitemShut {NoStop}%
\bibitem [{\citenamefont {Rissanen}(1978)}]{RISSANEN_1978}%
  \BibitemOpen
  \bibfield  {author} {\bibinfo {author} {\bibfnamefont {J.}~\bibnamefont
  {Rissanen}},\ }\bibfield  {title} {\bibinfo {title} {Modeling by shortest
  data description},\ }\href
  {https://doi.org/https://doi.org/10.1016/0005-1098(78)90005-5} {\bibfield
  {journal} {\bibinfo  {journal} {Automatica}\ }\textbf {\bibinfo {volume}
  {14}},\ \bibinfo {pages} {465} (\bibinfo {year} {1978})}\BibitemShut
  {NoStop}%
\bibitem [{\citenamefont {Balasubramanian}(1997)}]{Balasubramanian_1997}%
  \BibitemOpen
  \bibfield  {author} {\bibinfo {author} {\bibfnamefont {V.}~\bibnamefont
  {Balasubramanian}},\ }\bibfield  {title} {\bibinfo {title} {Statistical
  inference, occam's razor, and statistical mechanics on the space of
  probability distributions},\ }\href
  {https://doi.org/10.1162/neco.1997.9.2.349} {\bibfield  {journal} {\bibinfo
  {journal} {Neural Comput.}\ }\textbf {\bibinfo {volume} {9}},\ \bibinfo
  {pages} {349–368} (\bibinfo {year} {1997})}\BibitemShut {NoStop}%
\bibitem [{\citenamefont {Jeffreys}(1946)}]{Jeffreys_1946}%
  \BibitemOpen
  \bibfield  {author} {\bibinfo {author} {\bibfnamefont {H.}~\bibnamefont
  {Jeffreys}},\ }\bibfield  {title} {\bibinfo {title} {An invariant form for
  the prior probability in estimation problems},\ }\href
  {https://doi.org/10.1098/rspa.1946.0056} {\bibfield  {journal} {\bibinfo
  {journal} {Proceedings of the Royal Society of London. Series A. Mathematical
  and Physical Sciences}\ }\textbf {\bibinfo {volume} {186}},\ \bibinfo {pages}
  {453} (\bibinfo {year} {1946})}\BibitemShut {NoStop}%
\bibitem [{\citenamefont {Kelly~Jr.}(1956)}]{Kelly_1956}%
  \BibitemOpen
  \bibfield  {author} {\bibinfo {author} {\bibfnamefont {J.~L.}\ \bibnamefont
  {Kelly~Jr.}},\ }\bibfield  {title} {\bibinfo {title} {A new interpretation of
  information rate},\ }\href
  {https://doi.org/10.1002/j.1538-7305.1956.tb03809.x} {\bibfield  {journal}
  {\bibinfo  {journal} {Bell System Technical Journal}\ }\textbf {\bibinfo
  {volume} {35}},\ \bibinfo {pages} {917} (\bibinfo {year} {1956})}\BibitemShut
  {NoStop}%
\bibitem [{\citenamefont {Rivoire}\ and\ \citenamefont
  {Leibler}(2011)}]{rivoire2011value}%
  \BibitemOpen
  \bibfield  {author} {\bibinfo {author} {\bibfnamefont {O.}~\bibnamefont
  {Rivoire}}\ and\ \bibinfo {author} {\bibfnamefont {S.}~\bibnamefont
  {Leibler}},\ }\bibfield  {title} {\bibinfo {title} {The value of information
  for populations in varying environments},\ }\href@noop {} {\bibfield
  {journal} {\bibinfo  {journal} {Journal of Statistical Physics}\ }\textbf
  {\bibinfo {volume} {142}},\ \bibinfo {pages} {1124} (\bibinfo {year}
  {2011})}\BibitemShut {NoStop}%
\bibitem [{\citenamefont {Fisher}\ and\ \citenamefont
  {Russell}(1922)}]{Fisher_etal_1922}%
  \BibitemOpen
  \bibfield  {author} {\bibinfo {author} {\bibfnamefont {R.~A.}\ \bibnamefont
  {Fisher}}\ and\ \bibinfo {author} {\bibfnamefont {E.~J.}\ \bibnamefont
  {Russell}},\ }\bibfield  {title} {\bibinfo {title} {On the mathematical
  foundations of theoretical statistics},\ }\href
  {https://doi.org/10.1098/rsta.1922.0009} {\bibfield  {journal} {\bibinfo
  {journal} {Philosophical Transactions of the Royal Society of London. Series
  A, Containing Papers of a Mathematical or Physical Character}\ }\textbf
  {\bibinfo {volume} {222}},\ \bibinfo {pages} {309} (\bibinfo {year}
  {1922})}\BibitemShut {NoStop}%
\bibitem [{\citenamefont {Amari}(1983)}]{Amari_1983}%
  \BibitemOpen
  \bibfield  {author} {\bibinfo {author} {\bibfnamefont {S.-I.}\ \bibnamefont
  {Amari}},\ }\bibfield  {title} {\bibinfo {title} {A foundation of information
  geometry},\ }\href {https://doi.org/https://doi.org/10.1002/ecja.4400660602}
  {\bibfield  {journal} {\bibinfo  {journal} {Electronics and Communications in
  Japan (Part I: Communications)}\ }\textbf {\bibinfo {volume} {66}},\ \bibinfo
  {pages} {1} (\bibinfo {year} {1983})}\BibitemShut {NoStop}%
\bibitem [{\citenamefont {Nielsen}\ and\ \citenamefont
  {Garcia}(2011)}]{nielsen2011statistical}%
  \BibitemOpen
  \bibfield  {author} {\bibinfo {author} {\bibfnamefont {F.}~\bibnamefont
  {Nielsen}}\ and\ \bibinfo {author} {\bibfnamefont {V.}~\bibnamefont
  {Garcia}},\ }\href@noop {} {\bibinfo {title} {Statistical exponential
  families: A digest with flash cards}} (\bibinfo {year} {2011}),\ \Eprint
  {https://arxiv.org/abs/0911.4863} {arXiv:0911.4863 [cs.LG]} \BibitemShut
  {NoStop}%
\bibitem [{\citenamefont {Wainwright}\ \emph {et~al.}(2008)\citenamefont
  {Wainwright}, \citenamefont {Jordan} \emph
  {et~al.}}]{wainwright2008graphical}%
  \BibitemOpen
  \bibfield  {author} {\bibinfo {author} {\bibfnamefont {M.~J.}\ \bibnamefont
  {Wainwright}}, \bibinfo {author} {\bibfnamefont {M.~I.}\ \bibnamefont
  {Jordan}}, \emph {et~al.},\ }\bibfield  {title} {\bibinfo {title} {Graphical
  models, exponential families, and variational inference},\ }\href@noop {}
  {\bibfield  {journal} {\bibinfo  {journal} {Foundations and Trends in Machine
  Learning}\ }\textbf {\bibinfo {volume} {1}},\ \bibinfo {pages} {1} (\bibinfo
  {year} {2008})}\BibitemShut {NoStop}%
\bibitem [{\citenamefont {Bregman}(1967)}]{bregman1967relaxation}%
  \BibitemOpen
  \bibfield  {author} {\bibinfo {author} {\bibfnamefont {L.~M.}\ \bibnamefont
  {Bregman}},\ }\bibfield  {title} {\bibinfo {title} {The relaxation method of
  finding the common point of convex sets and its application to the solution
  of problems in convex programming},\ }\href@noop {} {\bibfield  {journal}
  {\bibinfo  {journal} {USSR computational mathematics and mathematical
  physics}\ }\textbf {\bibinfo {volume} {7}},\ \bibinfo {pages} {200} (\bibinfo
  {year} {1967})}\BibitemShut {NoStop}%
\bibitem [{\citenamefont {Nielsen}(2020)}]{nielsen2020elementary}%
  \BibitemOpen
  \bibfield  {author} {\bibinfo {author} {\bibfnamefont {F.}~\bibnamefont
  {Nielsen}},\ }\bibfield  {title} {\bibinfo {title} {An elementary
  introduction to information geometry},\ }\href@noop {} {\bibfield  {journal}
  {\bibinfo  {journal} {Entropy}\ }\textbf {\bibinfo {volume} {22}},\ \bibinfo
  {pages} {1100} (\bibinfo {year} {2020})}\BibitemShut {NoStop}%
\bibitem [{\citenamefont {Banerjee}\ \emph {et~al.}(2005)\citenamefont
  {Banerjee}, \citenamefont {Merugu}, \citenamefont {Dhillon}, \citenamefont
  {Ghosh},\ and\ \citenamefont {Lafferty}}]{banerjee2005clustering}%
  \BibitemOpen
  \bibfield  {author} {\bibinfo {author} {\bibfnamefont {A.}~\bibnamefont
  {Banerjee}}, \bibinfo {author} {\bibfnamefont {S.}~\bibnamefont {Merugu}},
  \bibinfo {author} {\bibfnamefont {I.~S.}\ \bibnamefont {Dhillon}}, \bibinfo
  {author} {\bibfnamefont {J.}~\bibnamefont {Ghosh}},\ and\ \bibinfo {author}
  {\bibfnamefont {J.}~\bibnamefont {Lafferty}},\ }\bibfield  {title} {\bibinfo
  {title} {Clustering with {B}regman {D}ivergences.},\ }\href@noop {}
  {\bibfield  {journal} {\bibinfo  {journal} {Journal of machine learning
  research}\ }\textbf {\bibinfo {volume} {6}} (\bibinfo {year}
  {2005})}\BibitemShut {NoStop}%
\bibitem [{\citenamefont {Diaconis}\ and\ \citenamefont
  {Ylvisaker}(1979)}]{DiaconisYlvisaker_1979}%
  \BibitemOpen
  \bibfield  {author} {\bibinfo {author} {\bibfnamefont {P.}~\bibnamefont
  {Diaconis}}\ and\ \bibinfo {author} {\bibfnamefont {D.}~\bibnamefont
  {Ylvisaker}},\ }\bibfield  {title} {\bibinfo {title} {Conjugate priors for
  exponential families},\ }\href@noop {} {\bibfield  {journal} {\bibinfo
  {journal} {The Annals of Statistics}\ }\textbf {\bibinfo {volume} {7}},\
  \bibinfo {pages} {269} (\bibinfo {year} {1979})}\BibitemShut {NoStop}%
\bibitem [{\citenamefont {Lidstone}(1920)}]{Lidstone_1920}%
  \BibitemOpen
  \bibfield  {author} {\bibinfo {author} {\bibfnamefont {G.~J.}\ \bibnamefont
  {Lidstone}},\ }\bibfield  {title} {\bibinfo {title} {Note on the general case
  of the bayes-laplace formula for inductive or a posteriori probabilities},\
  }\href@noop {} {\bibfield  {journal} {\bibinfo  {journal} {Transactions of
  the Faculty of Actuaries}\ }\textbf {\bibinfo {volume} {8}},\ \bibinfo
  {pages} {13} (\bibinfo {year} {1920})}\BibitemShut {NoStop}%
\bibitem [{\citenamefont {Nemenman}\ \emph {et~al.}(2001)\citenamefont
  {Nemenman}, \citenamefont {Shafee},\ and\ \citenamefont
  {Bialek}}]{Nemenman_etal2001}%
  \BibitemOpen
  \bibfield  {author} {\bibinfo {author} {\bibfnamefont {I.}~\bibnamefont
  {Nemenman}}, \bibinfo {author} {\bibfnamefont {F.}~\bibnamefont {Shafee}},\
  and\ \bibinfo {author} {\bibfnamefont {W.}~\bibnamefont {Bialek}},\
  }\bibfield  {title} {\bibinfo {title} {Entropy and inference, revisited},\
  }in\ \href@noop {} {\emph {\bibinfo {booktitle} {Proceedings of the 14th
  International Conference on Neural Information Processing Systems: Natural
  and Synthetic}}},\ \bibinfo {series and number} {NIPS’01}\ (\bibinfo
  {publisher} {MIT Press},\ \bibinfo {address} {Cambridge, MA, USA},\ \bibinfo
  {year} {2001})\ p.\ \bibinfo {pages} {471–478}\BibitemShut {NoStop}%
\bibitem [{\citenamefont {Mayer}\ \emph {et~al.}(2015)\citenamefont {Mayer},
  \citenamefont {Balasubramanian}, \citenamefont {Mora},\ and\ \citenamefont
  {Walczak}}]{Mayer5950}%
  \BibitemOpen
  \bibfield  {author} {\bibinfo {author} {\bibfnamefont {A.}~\bibnamefont
  {Mayer}}, \bibinfo {author} {\bibfnamefont {V.}~\bibnamefont
  {Balasubramanian}}, \bibinfo {author} {\bibfnamefont {T.}~\bibnamefont
  {Mora}},\ and\ \bibinfo {author} {\bibfnamefont {A.~M.}\ \bibnamefont
  {Walczak}},\ }\bibfield  {title} {\bibinfo {title} {How a well-adapted immune
  system is organized},\ }\href {https://doi.org/10.1073/pnas.1421827112}
  {\bibfield  {journal} {\bibinfo  {journal} {Proceedings of the National
  Academy of Sciences}\ }\textbf {\bibinfo {volume} {112}},\ \bibinfo {pages}
  {5950} (\bibinfo {year} {2015})}\BibitemShut {NoStop}%
\bibitem [{\citenamefont {Gremer}\ and\ \citenamefont
  {Venable}(2014)}]{Gremer_2014}%
  \BibitemOpen
  \bibfield  {author} {\bibinfo {author} {\bibfnamefont {J.~R.}\ \bibnamefont
  {Gremer}}\ and\ \bibinfo {author} {\bibfnamefont {D.~L.}\ \bibnamefont
  {Venable}},\ }\bibfield  {title} {\bibinfo {title} {Bet hedging in desert
  winter annual plants: optimal germination strategies in a variable
  environment.},\ }\href@noop {} {\bibfield  {journal} {\bibinfo  {journal}
  {Ecology letters}\ }\textbf {\bibinfo {volume} {17 3}},\ \bibinfo {pages}
  {380} (\bibinfo {year} {2014})}\BibitemShut {NoStop}%
\bibitem [{\citenamefont {Balaban}\ \emph {et~al.}(2004)\citenamefont
  {Balaban}, \citenamefont {Merrin}, \citenamefont {Chait}, \citenamefont
  {Kowalik},\ and\ \citenamefont {Leibler}}]{Balaban_etal_2004}%
  \BibitemOpen
  \bibfield  {author} {\bibinfo {author} {\bibfnamefont {N.}~\bibnamefont
  {Balaban}}, \bibinfo {author} {\bibfnamefont {J.}~\bibnamefont {Merrin}},
  \bibinfo {author} {\bibfnamefont {R.}~\bibnamefont {Chait}}, \bibinfo
  {author} {\bibfnamefont {L.}~\bibnamefont {Kowalik}},\ and\ \bibinfo {author}
  {\bibfnamefont {S.}~\bibnamefont {Leibler}},\ }\bibfield  {title} {\bibinfo
  {title} {Bacterial persistence as a phenotypic switch},\ }\href@noop {}
  {\bibfield  {journal} {\bibinfo  {journal} {Science}\ }\textbf {\bibinfo
  {volume} {305}},\ \bibinfo {pages} {1622 } (\bibinfo {year}
  {2004})}\BibitemShut {NoStop}%
\bibitem [{\citenamefont {Moyed}\ and\ \citenamefont
  {Bertrand}(1983)}]{Moyed768}%
  \BibitemOpen
  \bibfield  {author} {\bibinfo {author} {\bibfnamefont {H.~S.}\ \bibnamefont
  {Moyed}}\ and\ \bibinfo {author} {\bibfnamefont {K.~P.}\ \bibnamefont
  {Bertrand}},\ }\bibfield  {title} {\bibinfo {title} {hipa, a newly recognized
  gene of escherichia coli k-12 that affects frequency of persistence after
  inhibition of murein synthesis.},\ }\href@noop {} {\bibfield  {journal}
  {\bibinfo  {journal} {Journal of Bacteriology}\ }\textbf {\bibinfo {volume}
  {155}},\ \bibinfo {pages} {768} (\bibinfo {year} {1983})}\BibitemShut
  {NoStop}%
\bibitem [{\citenamefont {Mattingly}\ \emph {et~al.}(2021)\citenamefont
  {Mattingly}, \citenamefont {Kamino}, \citenamefont {Machta},\ and\
  \citenamefont {Emonet}}]{Mattingly2021.02.22.432091}%
  \BibitemOpen
  \bibfield  {author} {\bibinfo {author} {\bibfnamefont {H.}~\bibnamefont
  {Mattingly}}, \bibinfo {author} {\bibfnamefont {K.}~\bibnamefont {Kamino}},
  \bibinfo {author} {\bibfnamefont {B.}~\bibnamefont {Machta}},\ and\ \bibinfo
  {author} {\bibfnamefont {T.}~\bibnamefont {Emonet}},\ }\bibfield  {title}
  {\bibinfo {title} {E. coli chemotaxis is information-limited},\ }\bibfield
  {journal} {\bibinfo  {journal} {bioRxiv}\ }\href
  {https://doi.org/10.1101/2021.02.22.432091} {10.1101/2021.02.22.432091}
  (\bibinfo {year} {2021})\BibitemShut {NoStop}%
\bibitem [{\citenamefont {Mattingly}\ \emph {et~al.}(2018)\citenamefont
  {Mattingly}, \citenamefont {Transtrum}, \citenamefont {Abbott},\ and\
  \citenamefont {Machta}}]{Mattingly1760}%
  \BibitemOpen
  \bibfield  {author} {\bibinfo {author} {\bibfnamefont {H.~H.}\ \bibnamefont
  {Mattingly}}, \bibinfo {author} {\bibfnamefont {M.~K.}\ \bibnamefont
  {Transtrum}}, \bibinfo {author} {\bibfnamefont {M.~C.}\ \bibnamefont
  {Abbott}},\ and\ \bibinfo {author} {\bibfnamefont {B.~B.}\ \bibnamefont
  {Machta}},\ }\bibfield  {title} {\bibinfo {title} {Maximizing the information
  learned from finite data selects a simple model},\ }\href
  {https://doi.org/10.1073/pnas.1715306115} {\bibfield  {journal} {\bibinfo
  {journal} {Proceedings of the National Academy of Sciences}\ }\textbf
  {\bibinfo {volume} {115}},\ \bibinfo {pages} {1760} (\bibinfo {year}
  {2018})}\BibitemShut {NoStop}%
\bibitem [{\citenamefont {Quinn}\ \emph {et~al.}(2021)\citenamefont {Quinn},
  \citenamefont {Abbott}, \citenamefont {Transtrum}, \citenamefont {Machta},\
  and\ \citenamefont {Sethna}}]{quinn2021information}%
  \BibitemOpen
  \bibfield  {author} {\bibinfo {author} {\bibfnamefont {K.~N.}\ \bibnamefont
  {Quinn}}, \bibinfo {author} {\bibfnamefont {M.~C.}\ \bibnamefont {Abbott}},
  \bibinfo {author} {\bibfnamefont {M.~K.}\ \bibnamefont {Transtrum}}, \bibinfo
  {author} {\bibfnamefont {B.~B.}\ \bibnamefont {Machta}},\ and\ \bibinfo
  {author} {\bibfnamefont {J.~P.}\ \bibnamefont {Sethna}},\ }\href@noop {}
  {\bibinfo {title} {Information geometry for multiparameter models: New
  perspectives on the origin of simplicity}} (\bibinfo {year} {2021}),\ \Eprint
  {https://arxiv.org/abs/2111.07176} {arXiv:2111.07176 [cond-mat.stat-mech]}
  \BibitemShut {NoStop}%
\bibitem [{\citenamefont {Gutenkunst}\ \emph {et~al.}(2007)\citenamefont
  {Gutenkunst}, \citenamefont {Waterfall}, \citenamefont {Casey}, \citenamefont
  {Brown}, \citenamefont {Myers},\ and\ \citenamefont
  {Sethna}}]{gutenkunst2007universally}%
  \BibitemOpen
  \bibfield  {author} {\bibinfo {author} {\bibfnamefont {R.~N.}\ \bibnamefont
  {Gutenkunst}}, \bibinfo {author} {\bibfnamefont {J.~J.}\ \bibnamefont
  {Waterfall}}, \bibinfo {author} {\bibfnamefont {F.~P.}\ \bibnamefont
  {Casey}}, \bibinfo {author} {\bibfnamefont {K.~S.}\ \bibnamefont {Brown}},
  \bibinfo {author} {\bibfnamefont {C.~R.}\ \bibnamefont {Myers}},\ and\
  \bibinfo {author} {\bibfnamefont {J.~P.}\ \bibnamefont {Sethna}},\ }\bibfield
   {title} {\bibinfo {title} {Universally sloppy parameter sensitivities in
  systems biology models},\ }\href@noop {} {\bibfield  {journal} {\bibinfo
  {journal} {PLoS computational biology}\ }\textbf {\bibinfo {volume} {3}},\
  \bibinfo {pages} {e189} (\bibinfo {year} {2007})}\BibitemShut {NoStop}%
\bibitem [{\citenamefont {Schnaack}\ \emph {et~al.}(2021)\citenamefont
  {Schnaack}, \citenamefont {Peliti},\ and\ \citenamefont
  {Nourmohammad}}]{schnaack2021risk}%
  \BibitemOpen
  \bibfield  {author} {\bibinfo {author} {\bibfnamefont {O.~H.}\ \bibnamefont
  {Schnaack}}, \bibinfo {author} {\bibfnamefont {L.}~\bibnamefont {Peliti}},\
  and\ \bibinfo {author} {\bibfnamefont {A.}~\bibnamefont {Nourmohammad}},\
  }\bibfield  {title} {\bibinfo {title} {Risk-utility tradeoff shapes memory
  strategies for evolving patterns},\ }\href@noop {} {\bibfield  {journal}
  {\bibinfo  {journal} {arXiv preprint arXiv:2110.15008}\ } (\bibinfo {year}
  {2021})}\BibitemShut {NoStop}%
\bibitem [{\citenamefont {Lee}\ \emph {et~al.}(2022)\citenamefont {Lee},
  \citenamefont {Flack},\ and\ \citenamefont {Krakauer}}]{lee2022outsourcing}%
  \BibitemOpen
  \bibfield  {author} {\bibinfo {author} {\bibfnamefont {E.~D.}\ \bibnamefont
  {Lee}}, \bibinfo {author} {\bibfnamefont {J.~C.}\ \bibnamefont {Flack}},\
  and\ \bibinfo {author} {\bibfnamefont {D.~C.}\ \bibnamefont {Krakauer}},\
  }\bibfield  {title} {\bibinfo {title} {Outsourcing memory through niche
  construction},\ }\href@noop {} {\bibfield  {journal} {\bibinfo  {journal}
  {bioRxiv}\ ,\ \bibinfo {pages} {2022}} (\bibinfo {year} {2022})}\BibitemShut
  {NoStop}%
\bibitem [{\citenamefont {Feldman}(2016)}]{feldman2016simplicity}%
  \BibitemOpen
  \bibfield  {author} {\bibinfo {author} {\bibfnamefont {J.}~\bibnamefont
  {Feldman}},\ }\bibfield  {title} {\bibinfo {title} {The simplicity principle
  in perception and cognition},\ }\href@noop {} {\bibfield  {journal} {\bibinfo
   {journal} {Wiley Interdisciplinary Reviews: Cognitive Science}\ }\textbf
  {\bibinfo {volume} {7}},\ \bibinfo {pages} {330} (\bibinfo {year}
  {2016})}\BibitemShut {NoStop}%
\bibitem [{\citenamefont {Koffka}(2013)}]{koffka2013principles}%
  \BibitemOpen
  \bibfield  {author} {\bibinfo {author} {\bibfnamefont {K.}~\bibnamefont
  {Koffka}},\ }\href@noop {} {\emph {\bibinfo {title} {Principles of Gestalt
  psychology}}},\ Vol.~\bibinfo {volume} {44}\ (\bibinfo  {publisher}
  {routledge},\ \bibinfo {year} {2013})\BibitemShut {NoStop}%
\bibitem [{\citenamefont {Chater}\ and\ \citenamefont
  {Vit{\'a}nyi}(2003)}]{chater2003simplicity}%
  \BibitemOpen
  \bibfield  {author} {\bibinfo {author} {\bibfnamefont {N.}~\bibnamefont
  {Chater}}\ and\ \bibinfo {author} {\bibfnamefont {P.}~\bibnamefont
  {Vit{\'a}nyi}},\ }\bibfield  {title} {\bibinfo {title} {Simplicity: a
  unifying principle in cognitive science?},\ }\href@noop {} {\bibfield
  {journal} {\bibinfo  {journal} {Trends in cognitive sciences}\ }\textbf
  {\bibinfo {volume} {7}},\ \bibinfo {pages} {19} (\bibinfo {year}
  {2003})}\BibitemShut {NoStop}%
\bibitem [{\citenamefont {Gershman}\ and\ \citenamefont
  {Niv}(2013)}]{gershman2013perceptual}%
  \BibitemOpen
  \bibfield  {author} {\bibinfo {author} {\bibfnamefont {S.~J.}\ \bibnamefont
  {Gershman}}\ and\ \bibinfo {author} {\bibfnamefont {Y.}~\bibnamefont {Niv}},\
  }\bibfield  {title} {\bibinfo {title} {Perceptual estimation obeys occam's
  razor},\ }\href@noop {} {\bibfield  {journal} {\bibinfo  {journal} {Frontiers
  in psychology}\ }\textbf {\bibinfo {volume} {4}},\ \bibinfo {pages} {623}
  (\bibinfo {year} {2013})}\BibitemShut {NoStop}%
\bibitem [{\citenamefont {Tavoni}\ \emph {et~al.}(2022)\citenamefont {Tavoni},
  \citenamefont {Doi}, \citenamefont {Pizzica}, \citenamefont
  {Balasubramanian},\ and\ \citenamefont {Gold}}]{tavoni2022human}%
  \BibitemOpen
  \bibfield  {author} {\bibinfo {author} {\bibfnamefont {G.}~\bibnamefont
  {Tavoni}}, \bibinfo {author} {\bibfnamefont {T.}~\bibnamefont {Doi}},
  \bibinfo {author} {\bibfnamefont {C.}~\bibnamefont {Pizzica}}, \bibinfo
  {author} {\bibfnamefont {V.}~\bibnamefont {Balasubramanian}},\ and\ \bibinfo
  {author} {\bibfnamefont {J.~I.}\ \bibnamefont {Gold}},\ }\bibfield  {title}
  {\bibinfo {title} {Human inference reflects a normative balance of complexity
  and accuracy},\ }\href@noop {} {\bibfield  {journal} {\bibinfo  {journal}
  {Nature Human Behaviour}\ }\textbf {\bibinfo {volume} {6}},\ \bibinfo {pages}
  {1153} (\bibinfo {year} {2022})}\BibitemShut {NoStop}%
\bibitem [{\citenamefont {Piasini}\ \emph {et~al.}(2023)\citenamefont
  {Piasini}, \citenamefont {Liu}, \citenamefont {Chaudhari}, \citenamefont
  {Balasubramanian},\ and\ \citenamefont {Gold}}]{piasini2023occam}%
  \BibitemOpen
  \bibfield  {author} {\bibinfo {author} {\bibfnamefont {E.}~\bibnamefont
  {Piasini}}, \bibinfo {author} {\bibfnamefont {S.}~\bibnamefont {Liu}},
  \bibinfo {author} {\bibfnamefont {P.}~\bibnamefont {Chaudhari}}, \bibinfo
  {author} {\bibfnamefont {V.}~\bibnamefont {Balasubramanian}},\ and\ \bibinfo
  {author} {\bibfnamefont {J.~I.}\ \bibnamefont {Gold}},\ }\bibfield  {title}
  {\bibinfo {title} {How occam's razor guides human decision-making},\
  }\href@noop {} {\bibfield  {journal} {\bibinfo  {journal} {bioRxiv}\ ,\
  \bibinfo {pages} {2023}} (\bibinfo {year} {2023})}\BibitemShut {NoStop}%
\bibitem [{\citenamefont {Rissanen}(1996)}]{rissanen1996fisher}%
  \BibitemOpen
  \bibfield  {author} {\bibinfo {author} {\bibfnamefont {J.~J.}\ \bibnamefont
  {Rissanen}},\ }\bibfield  {title} {\bibinfo {title} {Fisher information and
  stochastic complexity},\ }\href@noop {} {\bibfield  {journal} {\bibinfo
  {journal} {IEEE transactions on information theory}\ }\textbf {\bibinfo
  {volume} {42}},\ \bibinfo {pages} {40} (\bibinfo {year} {1996})}\BibitemShut
  {NoStop}%
\bibitem [{\citenamefont {Gr{\"u}nwald}(2007)}]{grunwald2007minimum}%
  \BibitemOpen
  \bibfield  {author} {\bibinfo {author} {\bibfnamefont {P.~D.}\ \bibnamefont
  {Gr{\"u}nwald}},\ }\href@noop {} {\emph {\bibinfo {title} {The minimum
  description length principle}}}\ (\bibinfo  {publisher} {MIT press},\
  \bibinfo {year} {2007})\BibitemShut {NoStop}%
\bibitem [{\citenamefont {Bialek}\ \emph {et~al.}(2001)\citenamefont {Bialek},
  \citenamefont {Nemenman},\ and\ \citenamefont
  {Tishby}}]{bialek2001predictability}%
  \BibitemOpen
  \bibfield  {author} {\bibinfo {author} {\bibfnamefont {W.}~\bibnamefont
  {Bialek}}, \bibinfo {author} {\bibfnamefont {I.}~\bibnamefont {Nemenman}},\
  and\ \bibinfo {author} {\bibfnamefont {N.}~\bibnamefont {Tishby}},\
  }\bibfield  {title} {\bibinfo {title} {Predictability, complexity, and
  learning},\ }\href@noop {} {\bibfield  {journal} {\bibinfo  {journal} {Neural
  computation}\ }\textbf {\bibinfo {volume} {13}},\ \bibinfo {pages} {2409}
  (\bibinfo {year} {2001})}\BibitemShut {NoStop}%
\bibitem [{\citenamefont {Sachdeva}\ \emph {et~al.}(2021)\citenamefont
  {Sachdeva}, \citenamefont {Mora}, \citenamefont {Walczak},\ and\
  \citenamefont {Palmer}}]{sachdeva2021optimal}%
  \BibitemOpen
  \bibfield  {author} {\bibinfo {author} {\bibfnamefont {V.}~\bibnamefont
  {Sachdeva}}, \bibinfo {author} {\bibfnamefont {T.}~\bibnamefont {Mora}},
  \bibinfo {author} {\bibfnamefont {A.~M.}\ \bibnamefont {Walczak}},\ and\
  \bibinfo {author} {\bibfnamefont {S.~E.}\ \bibnamefont {Palmer}},\ }\bibfield
   {title} {\bibinfo {title} {Optimal prediction with resource constraints
  using the information bottleneck},\ }\href@noop {} {\bibfield  {journal}
  {\bibinfo  {journal} {PLOS Computational Biology}\ }\textbf {\bibinfo
  {volume} {17}},\ \bibinfo {pages} {e1008743} (\bibinfo {year}
  {2021})}\BibitemShut {NoStop}%
\bibitem [{\citenamefont {Murphy}(2007)}]{KevinMurphy_Gaussian}%
  \BibitemOpen
  \bibfield  {author} {\bibinfo {author} {\bibfnamefont {K.~P.}\ \bibnamefont
  {Murphy}},\ }\href@noop {} {\bibinfo {title} {Conjugate bayesian analysis of
  the gaussian distribution}},\ \bibinfo {howpublished}
  {\url{https://www.cs.ubc.ca/~murphyk/Papers/bayesGauss.pdf}} (\bibinfo {year}
  {2007}),\ \bibinfo {note} {(version: 2023-04-13)}\BibitemShut {NoStop}%
\end{thebibliography}%

\end{document}